\documentclass[twoside]{article}
\usepackage{LaThuileFPSproa}
\usepackage{graphics}

\input boxedeps
\SetepsfEPSFSpecial
\HideDisplacementBoxes

\newcommand{\slj}[3]{\mbox{$^{#1}${\ifcase#2\or S\or 
         P\or D\or F\or G\fi}$_{#3}$}}

\newcommand{\kev}{\hbox{ keV}}
\newcommand{\mev}{\hbox{ MeV}}
\newcommand{\gev}{\hbox{ GeV}}
\newcommand{\fm}{\hbox{ fm}}

\newcommand{\cfrac}[2]{\textstyle{\frac{#1}{#2}}}
\newcommand{\jpsi}{\ensuremath{J\!/\!\psi}}

\def\ltap{\,\raisebox{-.4ex}{\rlap{$\sim$}} \raisebox{.4ex}{$<$}\,}
\def\gtap{\,\raisebox{-.4ex}{\rlap{$\sim$}} \raisebox{.4ex}{$>$}\,}
\begin{document}
\title{ 
  QUARKONIUM: NEW DEVELOPMENTS  }
\author{
 Chris Quigg       \\
  {\em Fermi National Accelerator Laboratory} \\
  {\em P.O. Box 500, Batavia, Illinois 60510 USA}
  }
\maketitle

\baselineskip=11.6pt

\begin{abstract}
To illustrate the campaign to understand heavy quarkonium systems, I
focus on a puzzling new state, $X(3872) \to \pi^+\pi^-\jpsi$.  Studying
the influence of open-charm channels on charmonium properties leads us
to propose a new charmonium spectroscopy: additional discrete
charmonium levels that can be discovered as narrow resonances of
charmed and anticharmed mesons.  I recall some expectations for a new
spectroscopy of mesons with beauty and charm, the $B_c$ ($b\bar{c}$)
system.  Throughout, I call attention to open issues for theory and
experiment.
\end{abstract}
\newpage
\section{The Renaissance in Hadron Spectroscopy}
We live in exciting times for hadron spectroscopy. Over the past two years, experiments have uncovered a number of new narrow states that extend our knowledge of hadrons and challenge our understanding of the strong interaction. First came the discovery in the Belle experiment of $\eta_{c}^{\prime}$ in exclusive $B \to K K_{S} K^{-}\pi^{+}$ decays\cite{Choi:2002na}.  CLEO\cite{Asner:2003wv}, BaBar\cite{Wagner:2003qb}, and Belle\cite{Abe:2003ja} have confirmed and refined the discovery of $\eta_{c}^{\prime}$ in $\gamma\gamma$ collisions, fixing its mass and width as $M(\eta_{c}^{\prime}) = 3637.7 \pm 4.4 \mev$ and $\Gamma(\eta_{c}^{\prime}) = 19 \pm 10\mev$\cite{Skwarnicki:2003wn}.  The unexpectedly narrow $D_{sJ}$ states discovered by Babar\cite{Aubert:2003fg}, CLEO\cite{Besson:2003cp}, and Belle\cite{Abe:2003jk} provided the next surprise. Evidence for $\Theta^+(1540)$, a baryon state with $K^+n$ quantum numbers that do not occur in the simple $qqq$ quark-model description of baryons, captured headlines around the world. And finally (for now!) comes the discovery by Belle\cite{Choi:2003ue} of $X(3872) \to \pi^+\pi^-\jpsi$, rapidly confirmed by CDF\cite{Acosta:2003zx} and D\O\cite{Dzerotemp,JesikLT}. Each of these new states raises questions of interpretation, and offers opportunities.

\begin{figure}[tbhp] 
\centerline{\BoxedEPSF{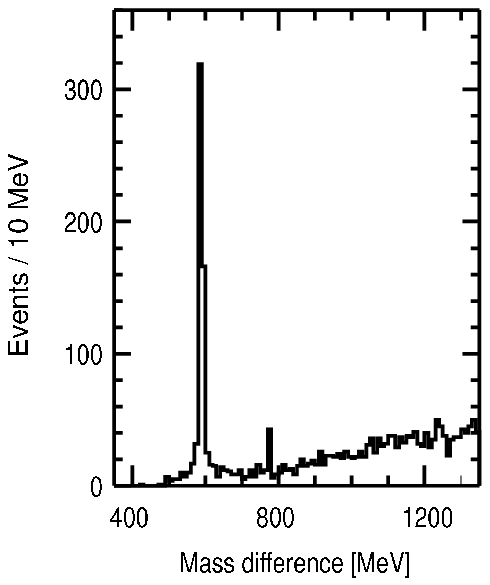 scaled 1200}}
\centerline{\BoxedEPSF{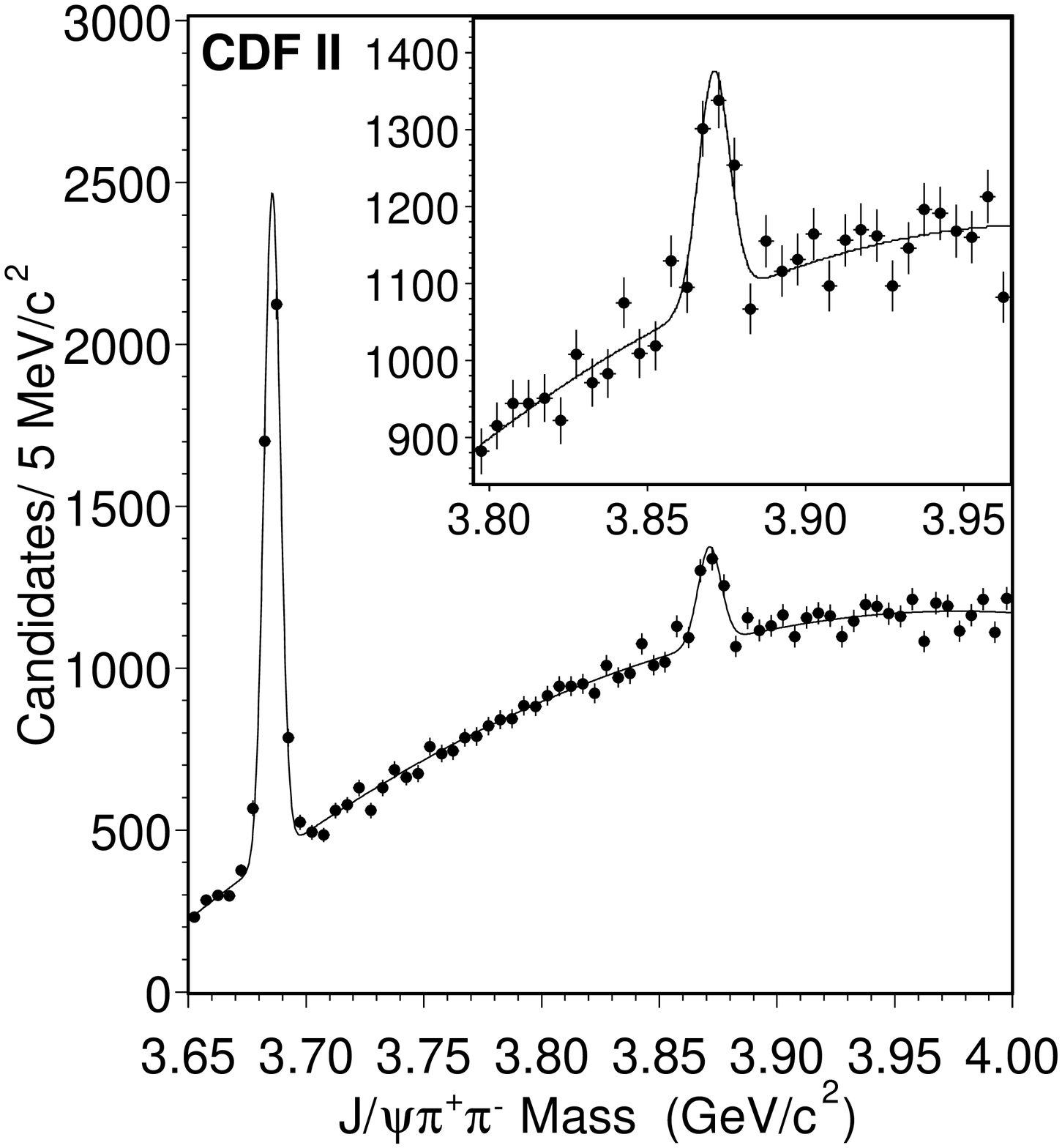 scaled 290} 
\BoxedEPSF{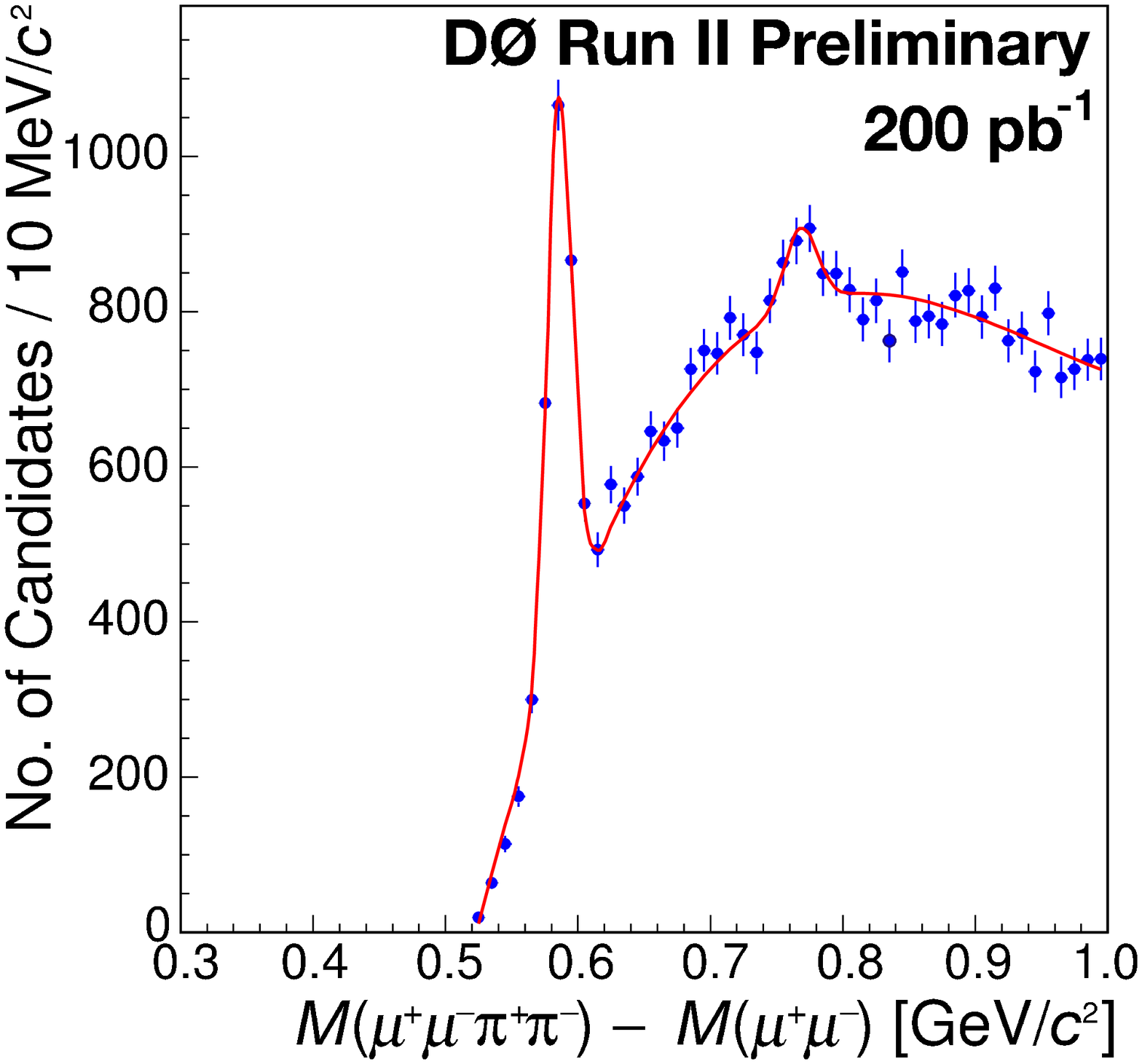 scaled 310}}
\vspace{-4pt}
\caption{Evidence for $X(3872) \to \pi^+\pi^-\jpsi$, from Belle\cite{Choi:2003ue}  (top panel), CDF\cite{Acosta:2003zx}  (bottom left), and D\O\cite{Dzerotemp}. (bottom right). The prominent peak on the left of each panel is $\psi^\prime(3686)$; the smaller peak near $\Delta M \equiv M(\pi^+\pi^-\ell^+\ell^-) - M(\ell^+\ell^-) \approx 775\mev, M(\jpsi\,\pi^+\pi^-)  \approx 3.87\gev$ is $X(3872)$. The CDF and D\O\ samples are restricted to dipion masses $>500$ and $520\mev$, respectively.}
\label{fig:X}
\end{figure}

The puzzle of $X(3872)$ will be the centerpiece of this talk, so I summarize the Belle, CDF, and D\O\ observations in Figure~\ref{fig:X}.

The outstanding issue for the \slj{1}{1}{0} $\eta_c^\prime$, compared to potential-model expectations, is the small splitting from its \slj{3}{1}{1} hyperfine partner $\psi^\prime$, which we shall examine presently. The $D_{sJ}(2317)$ and $D_{sJ}(2463)$ are apparently the  $0^{++}$ and $1^{++}$ levels of the $c\bar{s}$ system, corresponding to the $j_\ell = \cfrac{1}{2}$ doublet in the heavy-quark--symmetry classification. They surprised us by being lighter than their $j_\ell = \cfrac{3}{2}$ counterparts---so light that the expected strong decays into $KD^*$ are kinematically forbidden. Chiral symmetry\cite{Bardeen:2003kt} relates the $0^{++}$--$1^{++}$ doublet to the $0^{-+}$--$1^{--}$ ground-state doublet; we await detailed experimental tests. Assuming that $\Theta^+(1540)$ is confirmed, we need to learn the nature of this apparent pentaquark state. Is it best viewed as a chiral soliton, as uncorrelated $uudd\bar{s}$, or as correlated $[ud][ud]\bar{s}$ or other configurations involving diquarks?

Questions for $X(3872)$ include its mass, which differs from the simplest expectations for the \slj{3}{3}{2} charmonium state, and the nonobservation of radiative transitions. It is tantalizing that $X(3872)$ lies almost precisely at the $D^0\bar{D}^{*0}$ threshold. We will now take up the challenges of $X(3872)$ in detail.

\section{Charmonium in the Wake of the $\eta_c^\prime$ Discovery}
\begin{figure}[t!] 
\centerline{\BoxedEPSF{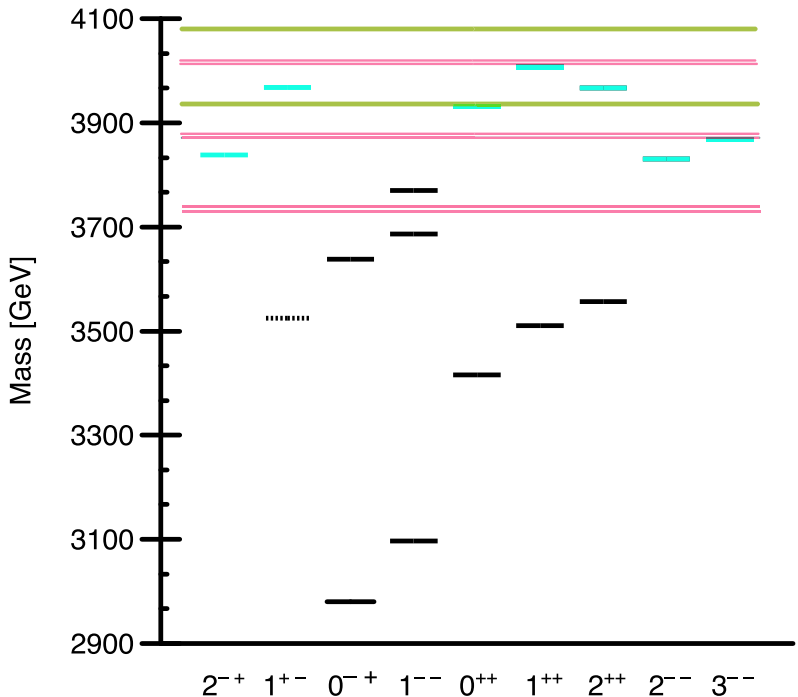 scaled 900}}
\vspace{-4pt}
\caption{Grotrian diagram for the charmonium spectrum. States marked by heavy black lines are well established. The \slj{1}{2}{1} $h_c$ level is indicated by the dashed line at the \slj{3}{2}{J} centroid. Thresholds are shown, in order of increasing mass, for $D^0\bar{D}^0$, $D^+D^-$, $D^0\bar{D}^{*0}$, $D^+\bar{D}^{*-}$, $D_s\bar{D}_s$,  $D^{*0}\bar{D}^{*0}$, $D^{*+}\bar{D}^{*-}$, and $D_s\bar{D}_s^*$. Some predicted states above threshold are depicted as faint lines.}
\label{fig:Grotrian}
\end{figure}

Charmonium is a fertile field that continues to draw our interest for many reasons\footnote{Vaia Papadimitriou's La Thuile talk\cite{VaiaLT} offers numerous concrete examples.}. Including the interthreshold region between $2M(D)$ and $M(D)+M(D^*)$, we expect about ten or eleven narrow levels, of which at least seven are already known. Including higher states within $800\mev$ of charm threshold, we expect perhaps sixty states, to be observed either as discrete levels or through their collective effect on the total cross section for $e^+e^- \to \hbox{  hadrons}$. A portion of the charmonium spectrum is shown in Figure~\ref{fig:Grotrian}.  Nonrelativistic potential models historically have given a good account of the spectrum, but they cannot be the whole story. They are truncated, single-channel treatments that do not contain the full richness of quantum chromodynamics. We are coming closer to a complete theoretical treatment: lattice QCD is increasingly capable for quarkonium spectroscopy---and improvements are coming swiftly. Charmonium states are being seen in electron-positron annihilations, in $B$ decay, in two-photon collisions, and in hadronic production. This circumstance gives us access to a very broad variety of quantum numbers $J^{PC}$, and makes for a lively conversation among experiments and a fruitful dialogue between theory and experiment.

Stimulated by Belle's discovery of $\eta_c^\prime$, my colleagues Estia Eichten, Ken Lane, and I sketched a coherent strategy  to explore 
$\eta_{c}^{\prime}$ and the remaining  charmonium states that do not 
decay into open charm,
$h_{c}(1\slj{1}{2}{1})$,
$\eta_{c2}(1\slj{1}{3}{2})$, and
$\psi_{2}(1\slj{3}{3}{2})$, through $B$-meson
gateways\cite{Eichten:2002qv}.  We argued that
 radiative transitions among charmonium levels and $\pi\pi$
cascades to lower-lying charmonia would enable the
identification of these states. Ko, Lee and Song\cite{Ko:1997rn} discussed
the observation of the narrow D states by photonic and pionic
transitions, and Suzuki\cite{Suzuki:2002sq} emphasized that the cascade
decay $B \rightarrow h_c K^{(*)} \rightarrow \gamma \eta_c K^{(*)}$
offers a promising technique to look for $h_{c}$. 

We noted that, according to current understanding of charmonium formation in $B$-decays\cite{Ko:1997rn,Bodwin:1992qr,Ko:1996iv,Yuan:1997we}, the states in question all should be produced at a level of $\approx\cfrac{1}{2}\%$. Moreover, the \slj{1,3}{3}{2} states should indeed be narrow, if their masses lie below $D\bar{D}^*$ threshold. Our 2002 estimates of the gluonic decay rates of all the 1D states, and of their $\pi\pi$ cascade rates to the charmonium ground state are given in Table~\ref{table:sdk}.
\begin{table}[b!]
\caption{Hadronic decay widths of charmonium 1D states in the single-channel potential model with 2002 inputs, from Ref.\cite{Eichten:2002qv}.\label{table:hadronic}}
\begin{center}
\begin{tabular}{|cccc|}
\hline
Level & Mass (MeV) & Transition & Partial Width (keV)  \\ \hline
1\slj{1}{3}{2} & 3815 & $\begin{array}{c} \eta_{c2} \to gg \\ \eta_{c2} \to \pi\pi\eta_c \end{array}$ & $\begin{array}{r} 110\kev \\ \approx 45\kev \end{array}$ \\[9pt]
1\slj{3}{3}{1} & 3770 & $\begin{array}{c} \psi \to ggg \\ \psi \to \pi\pi\jpsi \end{array}$ & $\begin{array}{r} 216\kev \\ 43 \pm 15\kev \end{array}$ \\[9pt]
1\slj{3}{3}{2} & 3815 & $\begin{array}{c} \psi_2 \to ggg \\ \psi_2 \to \pi\pi\jpsi \end{array}$ & $\begin{array}{r} 36\kev \\ \approx 45\kev \end{array}$ \\[9pt]
1\slj{3}{3}{3} & 3815 &   $\begin{array}{c} \psi_3 \to ggg  \\ \psi_3 \to \pi\pi\jpsi \end{array}$ & $\begin{array}{r} 102\kev \\ \approx 45\kev \end{array}$ \\
\hline
\end{tabular}
\end{center}
\label{table:sdk}
\end{table}
The annihilation rates were computed using standard expressions of perturbative QCD. We used the Wigner-Eckart theorem of the color-multipole expansion to set all the 1D cascade rates to a common value, normalized to an old estimate of the $\psi(3770) \to \pi\pi\jpsi$ decay rate. Both parts of this statement are weaknesses: the Wigner-Eckart relation  for E1-E1 transitions does not take into account kinematic differences that arise when the initial 1D states or the final 1S states are not degenerate in mass, and the normalizing rate is poorly known.  Here at La Thuile we have heard a final determination\cite{ZhuLT,Bai:2003hv}  from the current BES data set, $\mathcal{B}(1\slj{3}{3}{1} \to \pi^+\pi^-\jpsi) = (0.338 \pm 0.137 \pm 0.082)\%$, or $\Gamma(1\slj{3}{3}{1} \to \pi^+\pi^-\jpsi) = 80 \pm 32 \pm 21\kev$. This value is challenged  by a CLEO-$c$ limit\cite{Skwarnicki:2003wn}, $\mathcal{B}(1\slj{3}{3}{1} \to \pi^+\pi^-\jpsi) < 0.26\%$ at 90\% C.L. This is a terribly hard measurement, but a precise normalization for the 1D properties is urgently needed!

Although the $\pi\pi$ cascades promised plausible rates for the observation of $\eta_{c2}$ and $\psi_2$,  our estimates of the radiative (E1) transition rates were markedly larger. We computed, for example, $\Gamma(1\slj{3}{3}{2} \to \chi_{c2}\gamma) = 56\kev$,  $\Gamma(1\slj{3}{3}{2} \to \chi_{c1}\gamma) = 260\kev$. and $\Gamma(\slj{1}{3}{2} \to h_c\gamma) = 303\kev$. Normalizing to the 45-keV $\pi\pi$ cascade rate, we anticipated that $\mathcal{B}(h_c \to \eta_c\gamma) \approx \cfrac{2}{5}$, $\mathcal{B}(\eta_{c2} \to h_c\gamma) \approx \cfrac{2}{3}$, and $\mathcal{B}(\psi_2 \to \chi_{c1,2}\gamma) \approx \cfrac{4}{5}$, of which $\mathcal{B}(\psi_2 \to \chi_{c1}\gamma) \approx \cfrac{2}{3}$.

\section{What We Know about $X(3872)$}
The $X(3872)$ did indeed show itself in a search for narrow charmonium states such as 1\slj{3}{3}{2}, but the observed mass of $3871.7 \pm 0.6\mev$ is considerably higher than the prediscovery expectation of $3815\mev$. CDF and D\O\ have not yet determined the prompt (as opposed to $B$-decay) fraction of $X$ production, but it is highly plausible\cite{VaiaLT,JesikLT} that prompt production is not negligible. 
Belle's discovery paper\cite{Choi:2003ue} compares the rates of $X$ and $\psi^\prime$ production in $B$ decays, 
\begin{equation}
    \frac{\mathcal{B}(B^{+} \!\to \!K^{+}X) 
    \mathcal{B}(X \!\to \!
    \pi^{+}\pi^{-}\jpsi)}{\mathcal{B}(B^{+}\!\to\! K^{+}\psi^{\prime}) 
    \mathcal{B}(\psi^{\prime}\!\to\! \pi^{+}\pi^{-}\jpsi)} = 
    0.063 \pm 0.014\; .
\end{equation}
Belle has searched in vain for radiative transitions to
the 1\slj{3}{2}{1} level; their 90\%\ C.L.\ upper bound\cite{Choi:2003ue,ChoiLL},
\begin{equation}
    \frac{\Gamma(X(3872) \to \gamma \chi_{c1})}{\Gamma(X(3872) \to 
    \pi^{+}\pi^{-}\jpsi)} < 0.89\; ,
    \label{eqn:radbound1}
\end{equation}
conflicts with our single-channel potential-model expectations
 for the 1\slj{3}{3}{2} state\cite{Eichten:2002qv}, while the limit\cite{ChoiLL}
\begin{equation}
    \frac{\Gamma(X(3872) \to \gamma \chi_{c2})}{\Gamma(X(3872) \to 
    \pi^{+}\pi^{-}\jpsi)} < 1.1\; ,
    \label{eqn:radbound2}
\end{equation}
is problematic for both the 1\slj{3}{3}{2} and 1\slj{3}{3}{3} interpretations.
The theoretical estimate of the $\pi\pi\jpsi$ rate is highly uncertain, however.

Just before we met in La Thuile, Belle\cite{ChoiLL} presented the first 
information about the decay angular distribution of $\jpsi$ produced in 
$X \to \pi^+\pi^-\jpsi$. It does not yet determine $J^{PC}$, 
but the 2\slj{1}{2}{1} $h_c^\prime$ (or $1^{+-}$ charm molecule) 
assignment is ruled out. For more on the diagnostic capabilities of decay angular distributions, see Jackson's classic Les Houches lectures\cite{JDJhouches} and the recent paper on $X(3872)$ by Pakvasa and Suzuki\cite{Pakvasa:2003ea}. A BES limit\cite{Yuan:2003yz} on the electronic width of $X(3872)$ argues against a $1^{--}$ assignment.

\section{Alternatives to Charmonium}
The notion that charm molecules might be formed by attractive pion
exchange between $D$ and $\bar{D}^*$ mesons has a long history, and has
been invoked as a possible interpretation for $X(3872)$ by
T\"{o}rnqvist\cite{Tornqvist:2004qy} and
others\cite{Voloshin:2003nt,Wong:2003xk,Swanson:2003tb}.  A maximally
attractive channel analysis suggests that deuteron-analogue
``deusons,'' as T\"{o}rnqvist likes to call them, should be $J^{PC} =
0^{-+}\hbox{ or }1^{++}$ states.  Parity conservation forbids the decay
of these levels into $(\pi\pi)_{I=0}\jpsi$; the isospin-violating
$(\pi\pi)_{I=1}\jpsi$ mode is required.  Although an isovector dipion
might account for the observed preference for high dipion masses, it
remains to be seen whether the decay rate is large enough.  (The
$D^+$-$D^0$ and $D^{*+}$-$D^{*0}$ mass splitting means that the
molecule is not a pure isoscalar state.)  T\"{o}rnqvist has suggested
that the dissociation $X(3872) \to (D^0\bar{D}^{*0})_{\mathrm{virtual}}
\to D^0 \bar{D}^0 \pi^0$ should be a prominent decay mode of a charm
molecule, with a partial width of perhaps $50\kev$.  The Belle
Collaboration's limit\cite{Abe:2003zv}, 
\begin{equation}
\mathcal{B}(B^+ \to K^+ X(3872))\,\mathcal{B}(X(3872) \to D^0 \bar{D}^0 \pi^0) < 6 \times 10^{-5}\;,
\end{equation}
is perhaps an order of magnitude from challenging this expectation. 

\begin{figure}[tb] 
\centerline{\BoxedEPSF{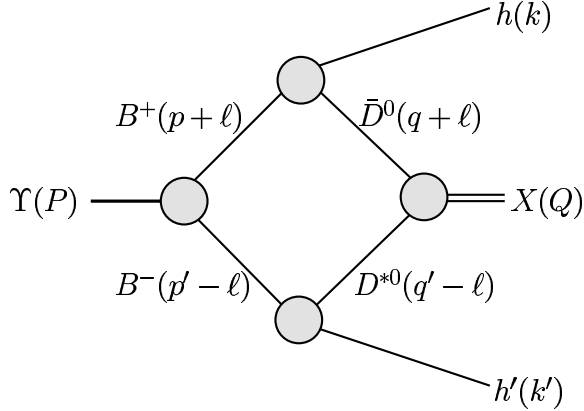 scaled 800}}
\vspace{-4pt}
\caption{Formation of charm molecule $X(3872)$ by coalescencing $\bar{D}^0$ and $D^{*0}$.}
\label{fig:BK}
\end{figure}

Braaten \& Kusunoki\cite{Braaten:2003he} conjecture that a charm
molecule that lies very close to threshold has universal properties
determined by an unnaturally large scattering length that is inversely
proportional to the reduced mass and the binding energy.  Both
production and decay rates would be suppressed by a factor of
$(\hbox{scattering length})^{-1}$.  The same
authors\cite{Braaten:2004rn} have calculated the probability for
charmed mesons produced in $\Upsilon(4\mathrm{S})$ decay to coalesce
into a lightly bound $D\bar{D}^*$ molecule by the mechanism shown in
Figure~\ref{fig:BK}.  The leading contribution is a universal form
proportional to $(\Gamma(B \to \hbox{all})/M_B)^2$ that depends only on
hadron masses and on the width and branching fractions of the $B$
meson, and on the binding energy $E_b$ of the molecule.  For light
binding, they find
\begin{equation}
\frac{\Gamma(\Upsilon(4\mathrm{S}) \to X(3872)hh^\prime)}{\Gamma(\Upsilon(4\mathrm{S})\to D^0\bar{D}^{*0}hh^\prime) + \Gamma(\Upsilon(4\mathrm{S})\to \bar{D}^0{D}^{*0}hh^\prime)} \approx 10^{-24}\; ,
\end{equation}
which may hold the distinction of being the smallest strong-interaction
branching fraction ever calculated!

Hybrid states such as $c\bar{c}g$ that manifest the gluonic degrees of freedom might also appear in the charmonium spectrum, and should be examined as interpretations of $X(3872)$\cite{Close:2003mb}. It is fair to say that dynamical calculations of hybrid-meson properties are in a primitive state, but heuristic arguments do offer some guidance. A picture based on chromoelectric flux tubes suggests that the lowest-lying states might have quantum numbers $J^{PC} = (0,1,2)^{++}\hbox{ or }1^{+-}$; for chromomagnetic flux tubes, $J^{PC} = (0,1,2)^{-+}\hbox{ or }1^{--}$ prevail. Ground-state $c\bar{c}g$ masses are anticipated around $4100 \pm 200\mev$, rather higher than $X(3872)$, but the estimates are not reliable enough to lead us immediately to dismiss the hybrid interpretation. The valence gluon in the hybrid wave function leads to the speculation that the $\eta\jpsi$ mode might be quite prominent. The Babar experiment\cite{Aubert:2004fc} has found no sign of $X \to \eta\jpsi$ and quoted a limit,
\begin{equation}
\mathcal{B}(X(3872) \to \eta\jpsi) < 2 \mathcal{B}(\psi^\prime \to \eta\jpsi)\;,
\end{equation}
that does not favor a privileged role for the $\eta\jpsi$ mode.

\section{Coupling to Open-Charm Channels}
The Cornell group showed long ago that a very simple model that
 couples charmonium to charmed-meson decay channels confirms the
 adequacy of the single-channel $c\bar{c}$ analysis below threshold and
 gives a qualitative understanding of the structures observed above
 threshold\cite{Eichten:1978tg,Eichten:1980ms}. Eichten and Lane and I recently 
have employed the 
 Cornell coupled-channel formalism to analyze the properties of 
 charmonium levels that populate the threshold region between $2M_D$ 
 and $2M_{D^*}$\cite{Eichten:2004uh}.

Our  command of quantum chromodynamics is inadequate to derive
 a realistic description of the interactions that communicate between
 the $c\bar{c}$ and $c\bar{q}+\bar{c}q$ sectors.  The Cornell formalism
 generalizes the $c\bar{c}$ model 
 without introducing new parameters, writing the interaction 
 Hamiltonian in second-quantized form as
 \begin{equation}
     \mathcal{H}_{I} = \cfrac{3}{8} \sum_{a=1}^{8} 
     \int:\rho_{a}(\mathbf{r}) V(\mathbf{r} - 
     \mathbf{r}^{\prime})\rho_{a}(\mathbf{r}^{\prime}): 
     d^{3}{r}\,d^{3}{r}^{\prime}\; ,
     \label{eq:CCCMH}
 \end{equation}
 where $V$ is the charmonium potential and $\rho_{a}(\mathbf{r}) = 
 \cfrac{1}{2}\psi^{\dagger}(\mathbf{r})\lambda_{a}\psi(\mathbf{r})$ is the color 
 current density, with $\psi$ the quark field operator and 
 $\lambda_{a}$ the octet of SU(3) matrices. To generate the relevant 
 interactions, $\psi$ is expanded in creation and annihilation 
 operators (for charm, up, down, and strange quarks), but transitions 
 from two mesons to three mesons and all transitions that violate the 
 Zweig rule are omitted. It is a good approximation to neglect all 
 effects of the Coulomb piece of the potential in (\ref{eq:CCCMH}).

 \begin{table}[tb]
 \caption{Charmonium spectrum, including the influence of open-charm 
 channels. All masses are in MeV. The penultimate column holds an estimate 
 of the spin splitting due to tensor and spin-orbit forces in a 
 single-channel potential model. The last 
 column gives the spin splitting induced by communication with 
 open-charm states, for an initially unsplit multiplet.
 \label{table:delM}}
 \begin{center}
 \begin{tabular}{|ccccc|}
 \hline 
 State & Mass & Centroid & $\begin{array}{c} \textrm{Splitting} 
 \\ \textrm{(Potential)} \end{array}$ &  $\begin{array}{c} \textrm{Splitting} 
 \\ \textrm{(Induced)} \end{array}$ \\
 \hline
 & & & & \\[-9pt]
 $\begin{array}{c}
     1\slj{1}{1}{0} \\
     1\slj{3}{1}{1} 
 \end{array}$ &
 $\begin{array}{c} 2\,979.9 \\ 3\,096.9 
 \end{array}$ & $3\,067.6$ & $\begin{array}{c} -90.5 
 \\ +30.2 \end{array}$
    &  $\begin{array}{c} +2.8  \\ -0.9 \end{array} $\\ & & & & \\[-6pt]
 $\begin{array}{c} 
 1\slj{3}{2}{0} \\
 1\slj{3}{2}{1} \\
 1\slj{1}{2}{1} \\
 1\slj{3}{2}{2}  \end{array}$ &  $\begin{array}{c}
 3\,415.3\\
 3\,510.5\\
 3\,525.3\\
 3\,556.2 \end{array}$
 & $3\,525.3$ &  $\begin{array}{c} 
 -114.9 \\ -11.6
 \\ +1.5 \\ -31.9 \end{array}$ & 
 $\begin{array}{c} +5.9 \\ -2.0 \\ +0.5 \\ -0.3
 \end{array}$ \\
 & & & & \\[-6pt]
 $\begin{array}{c}
     2\slj{1}{1}{0} \\
     2\slj{3}{1}{1} 
 \end{array}$ &
 $\begin{array}{c} 3\,637.7 \\ 3\,686.0
 \end{array}$ & $3\,673.9$ & $\begin{array}{c} -50.4 
 \\ +16.8 \end{array}$
    &  $\begin{array}{c} +15.7  \\ -5.2 \end{array} $\\
    & & & & \\[-6pt]
 $\begin{array}{c} 
 1\slj{3}{3}{1} \\
 1\slj{3}{3}{2} \\
 1\slj{1}{3}{2} \\
 1\slj{3}{3}{3}  \end{array}$ &  $\begin{array}{c}
 3\,769.9\\
 3\,830.6\\
 3\,838.0\\
 3\,868.3 \end{array}$
 & 
 (3\,815) & 
 $\begin{array}{c} -40 \\ 0 \\ 
 0 \\ +20 \end{array} $ & 
 $\begin{array}{c} -39.9 \\ -2.7\\ +4.2 \\ +19.0
 \end{array}$ \\ & & & & \\[-6pt]
 $\begin{array}{c} 
 2\slj{3}{2}{0} \\
 2\slj{3}{2}{1} \\
 2\slj{1}{2}{1} \\
 2\slj{3}{2}{2}  \end{array}$ &  $\begin{array}{c}
 3\,931.9\\
 4\,007.5\\
 3\,968.0\\
 3\,966.5 \end{array}$
 & 3\,968 & 
 $\begin{array}{c} -90 \\ -8 \\ 
 0 \\ +25 \end{array} $ & 
 $\begin{array}{c} +10 \\ +28.4 \\ -11.9 \\ -33.1
 \end{array}$ \\[3pt]
 \hline
 \end{tabular}
 \end{center}
 \end{table}
The basic coupled-channel interaction (\ref{eq:CCCMH}) is
 spin-independent, but the hyperfine splittings of $D$ and $D^{*}$, 
 $D_{s}$ and $D_{s}^{*}$, induce spin-dependent forces that affect the
 charmonium states.  These spin-dependent forces give rise to S-D
 mixing that contributes to the $\psi(3770)$ electronic width, for
 example, and are a source of additional spin splitting, shown in the
 rightmost column of Table~\ref{table:delM}.  To compute the induced 
 splittings, we adjust the bare centroid of the spin-triplet states so 
 that the physical centroid, after inclusion of coupled-channel 
 effects, matches the value in the middle column of 
 Table~\ref{table:delM}. 
 As expected, the shifts
 induced in the low-lying 1S and 1P levels are small.  For the other
 known states in the 2S and 1D families, coupled-channel effects are
 noticeable and interesting.  

In a simple potential picture, the $\eta_{c}^{\prime}(2\slj{1}{1}{0})$ 
 level lies below the $\psi^{\prime}(2\slj{3}{1}{1})$ by the hyperfine 
 splitting given by $M(\psi^{\prime}) - M(\eta_{c}^{\prime}) =
 32\pi\alpha_{s}|\Psi(0)|^{2}/9m_{c}^{2}$. Normalizing to the observed 
 1S hyperfine splitting, $M(\jpsi) - M(\eta_{c}) = 117\mev$, we  
 would find 
  \begin{equation}
     M(\psi^{\prime}) - M(\eta_{c}^{\prime}) = 67\mev\; ,
 \end{equation}
which is larger than the observed $48.3 \pm 4.4\mev$, as is typical 
for potential-model calculations.  The 2S induced
 shifts in Table~\ref{table:delM} draw $\psi^{\prime}$ and
 $\eta_{c}^{\prime}$ closer by $20.9\mev$, substantially improving the
 agreement between theory and experiment.  It is tempting to conclude 
 that the $\psi^{\prime}$-$\eta_{c}^{\prime}$ splitting reflects the 
 influence of virtual decay channels, but compare the analysis of 
 Ref.~\cite{Recksiegel:2003fm}.

We peg the 1D masses to the observed mass of the 1\slj{3}{3}{1} $\psi(3770)$. In our model calculation, the coupling to open-charm
 channels increases the 1\slj{3}{3}{2}-1\slj{3}{3}{1} splitting by
 about $20\mev$, but does not fully account for the observed $102\mev$
 separation between $X(3872)$ and $\psi(3770)$.  It is noteworthy that
 the position of the $3^{--}$ 1\slj{3}{3}{3} level turns out to be very
 close to $3872\mev$.  For the 2P levels, we have no experimental 
 anchor, so we adjust the bare centroid so that the 2\slj{1}{2}{1} 
 level lies at the centroid of the potential-model calculation. 

The physical charmonium states are not pure potential-model 
eigenstates. To compute the E1 radiative transition rates, we must 
take into account both the standard $(c\bar{c}) \to (c\bar{c})\gamma$ 
transitions and the transitions between (virtual) decay channels in 
the initial and final states. Our expectations for E1 decays of the 1\slj{3}{3}{2} and 1\slj{3}{3}{3} candidates for $X(3872)$ are
shown in Table~\ref{table:radtranstw}. 

\begin{table}[b]
\caption{Calculated rates for E1 radiative decays of some 1D levels. \textit{Values in italics} result if the influence of 
open-charm channels is not included.\label{table:radtranstw}}
\begin{center}
\begin{tabular}{|cc|} 
\hline
    Transition ($\gamma$ energy in MeV)  & Partial width (keV) \\
\hline
 &  \\[-6pt]
$1\slj{3}{3}{2}(3872)\to\chi_{c2}\,\gamma(303)$ & 
 $\mathit{85} \to 45$  \\
 $1\slj{3}{3}{2}(3872)\to\chi_{c1}\,\gamma(344)$ & 
  $\mathit{362} \to 207$  \\[6pt]
    $1\slj{3}{3}{3}(3872)\to\chi_{c2}\,\gamma(304)$ & 
     $\mathit{341} \to 299$  \\[3pt]
\hline
\end{tabular}
\end{center}
\end{table}

\begin{figure}
    \BoxedEPSF{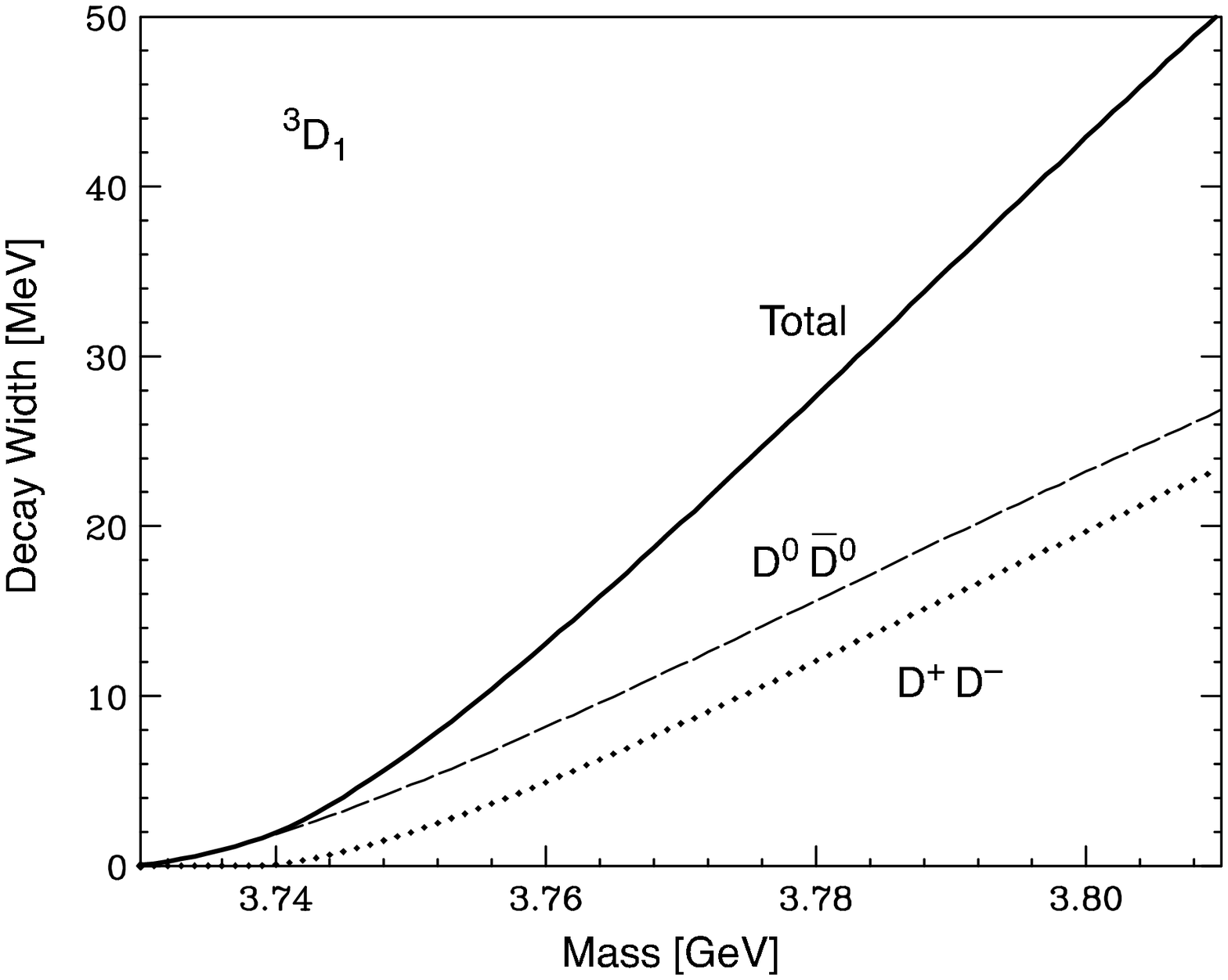 scaled 310}\quad
    \BoxedEPSF{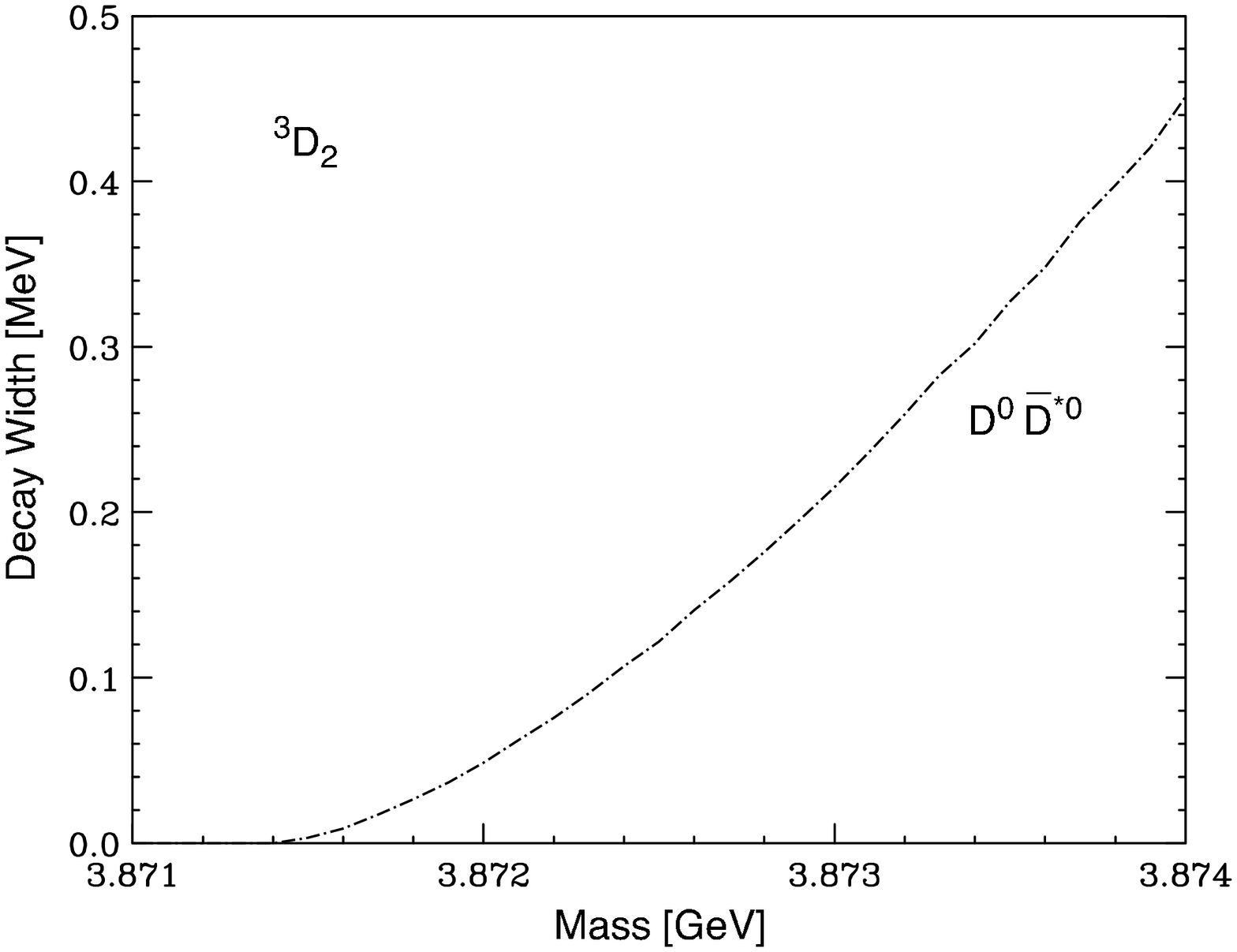 scaled 310}\\[3pt]
    \BoxedEPSF{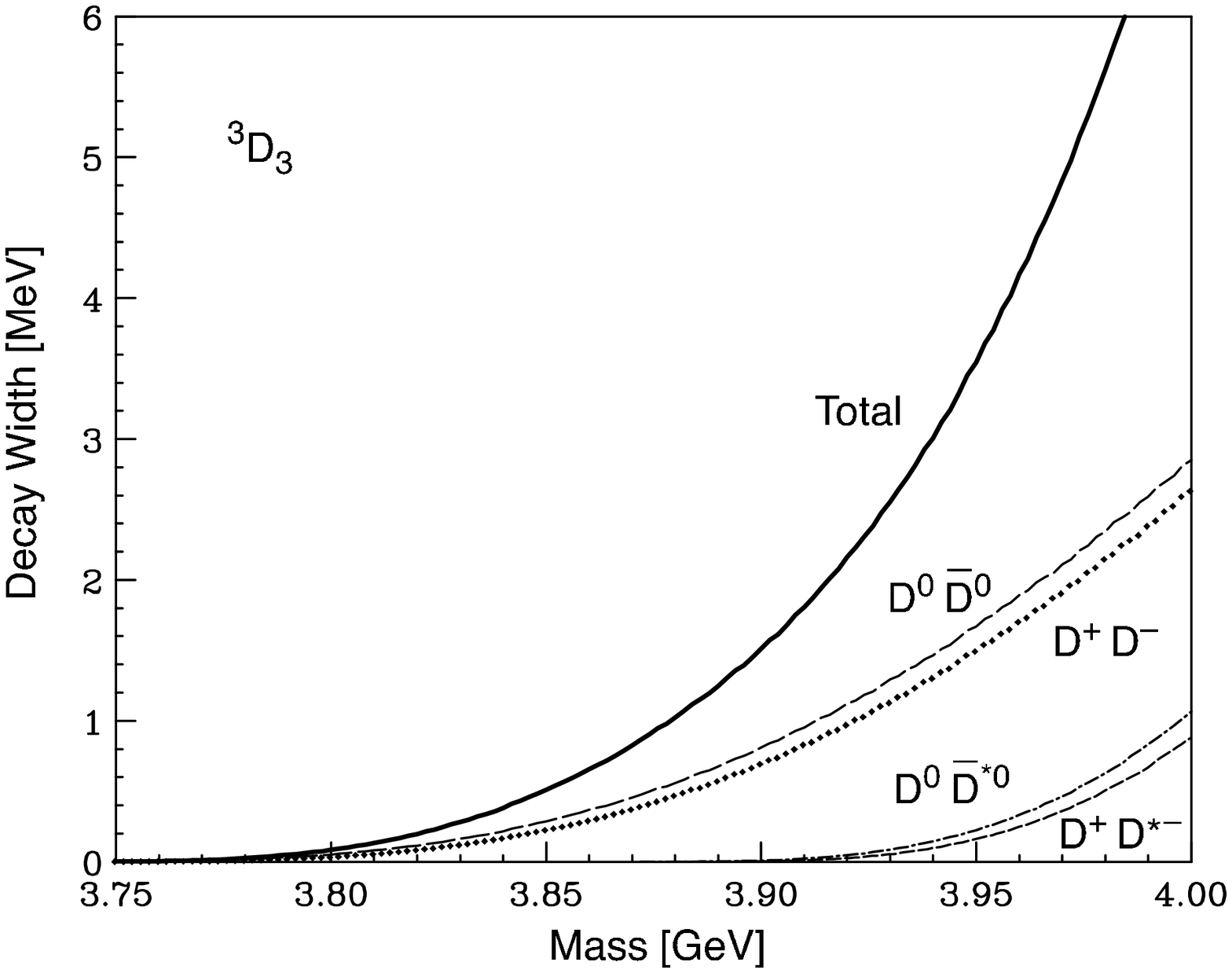 scaled 310}\quad
    \BoxedEPSF{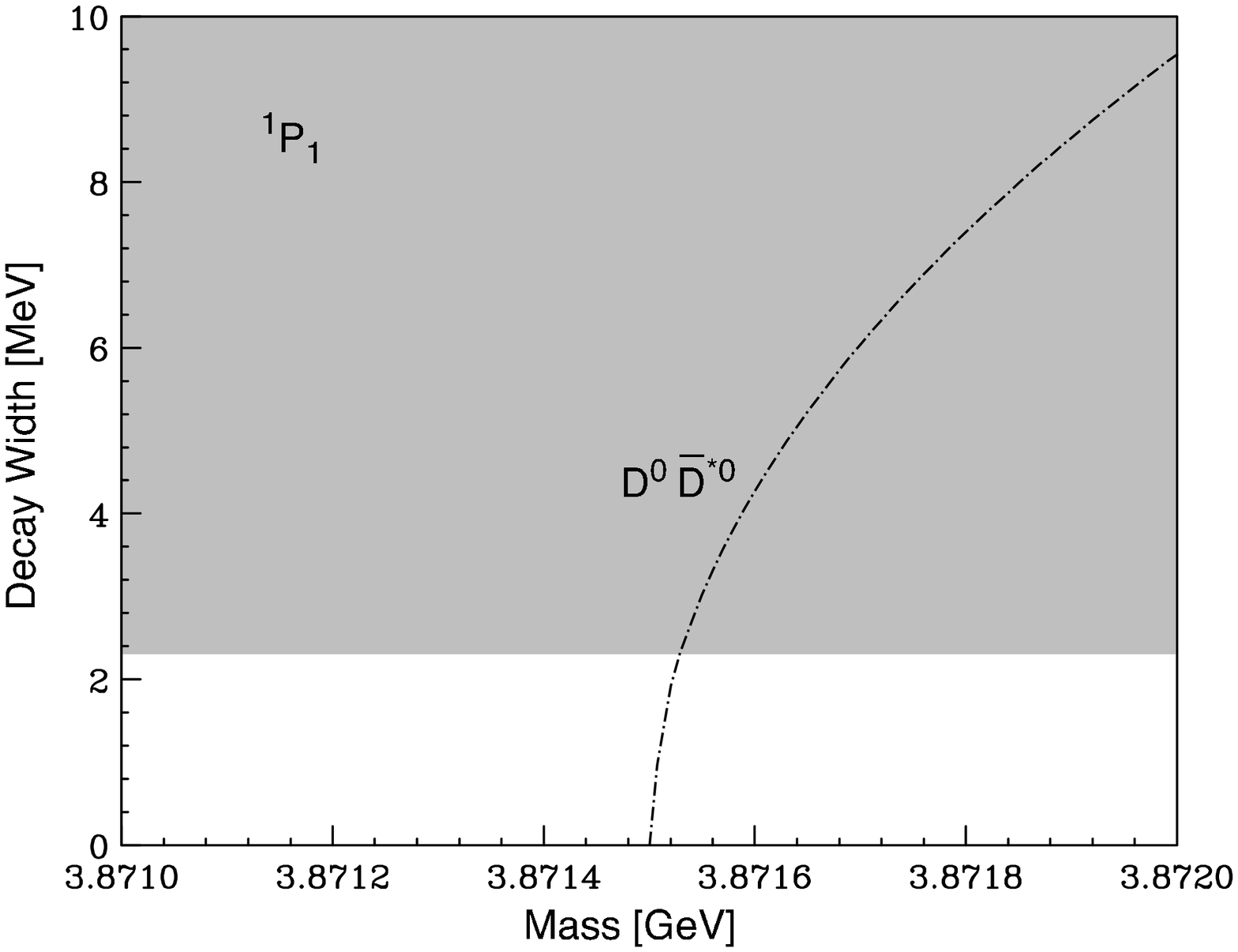 scaled 310}\\[3pt]
    \BoxedEPSF{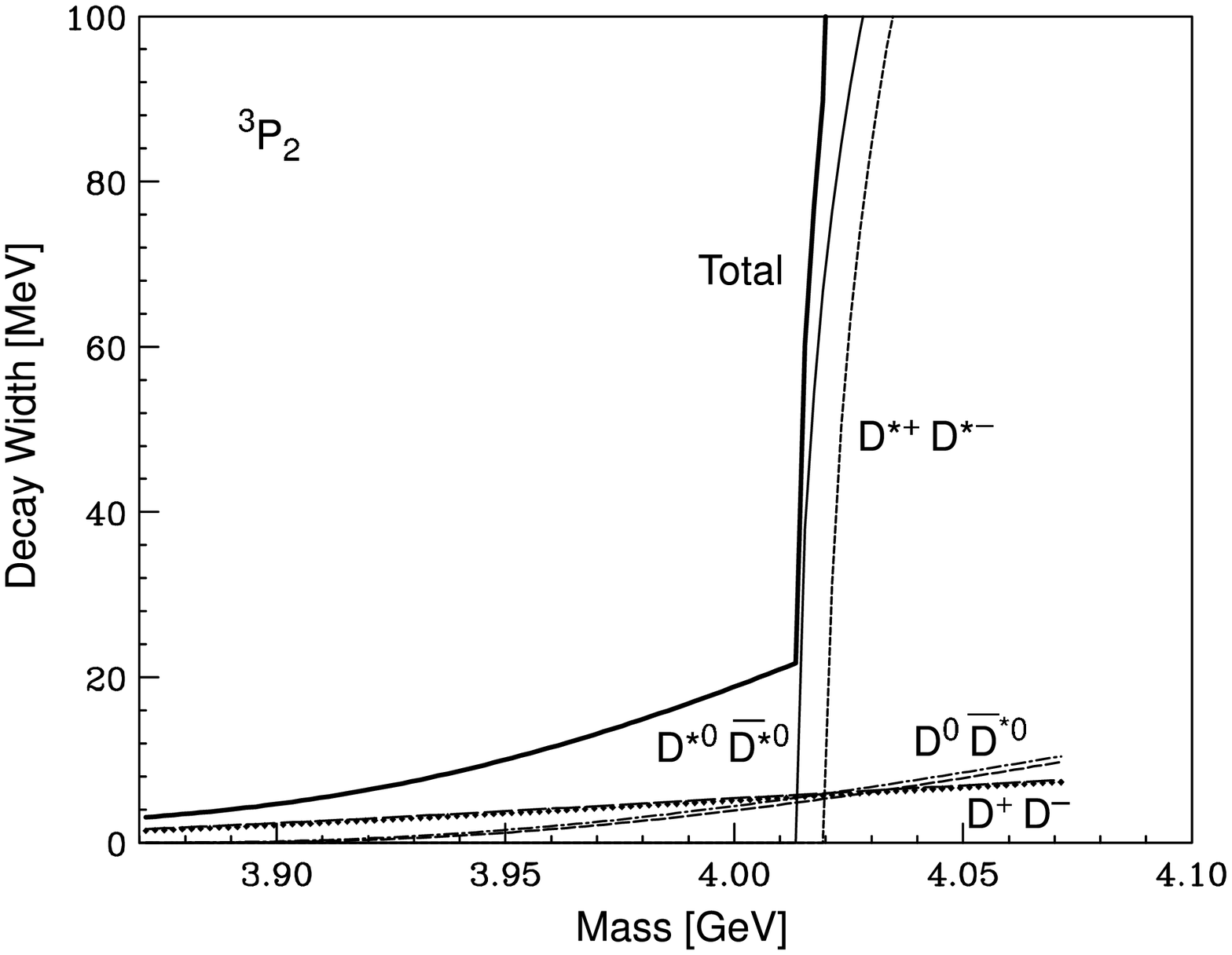 scaled 310}\quad
   \BoxedEPSF{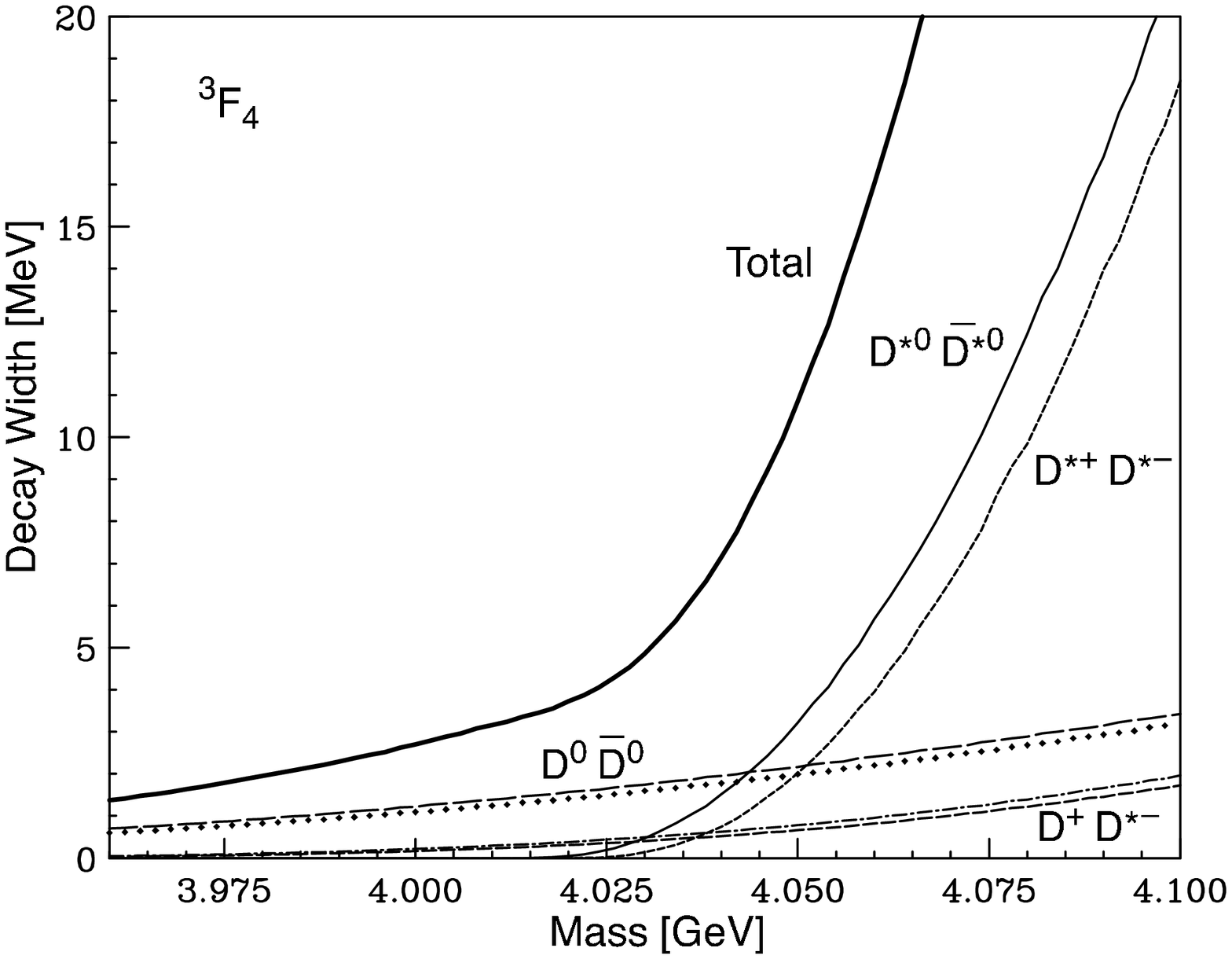 scaled 310}
    
    \caption{Partial and total widths near threshold for decay of 
    charmonium states into open charm, computed in the Cornell 
    coupled-channel model. Long dashes: $D^{0}\bar{D}^{0}$, dots: 
    $D^{+}D^{-}$, dot-dashes: $D^{0}\bar{D}^{*0}$, dashes: $D^{+}D^{*-}$, 
    thin line: $D^{*0}\bar{D}^{*0}$, short dashes: $D^{*+}D^{*-}$, 
    widely spaced dots: $D_{s}\bar{D}_{s}$,
    thick line: sum of open-charm channels. Belle's 90\% C.L.\ upper 
    limit\cite{Choi:2003ue}, $\Gamma(X(3872)) < 2.3\mev$, is 
    indicated on the \slj{1}{2}{1} window. For $D\bar{D}^{*}$ modes, 
    the sum of $D\bar{D}^{*}$ and $\bar{D}D^{*}$ is always implied.   \label{figure:OCdecays}}
\end{figure}

Once the position of a resonance is given, the coupled-channel 
formalism yields reasonable predictions for the other resonance 
properties. The 1\slj{3}{3}{1} state $\psi^{\prime\prime}(3770)$, 
which lies some $40\mev$ above charm threshold,  
offers an important benchmark: we compute 
$\Gamma( \psi^{\prime\prime}(3770)\to 
D\bar{D}) = 20.1\mev$, to be compared with the Particle Data Group's 
fitted value of $23.6 \pm 2.7\mev$\cite{PDBook}. The variation of the 
1\slj{3}{3}{1} width with mass is shown in the top left panel of 
Figure~\ref{figure:OCdecays}.  

Barnes \& Godfrey\cite{Barnes:2003vb} have estimated the decays of
several of the charmonium states into open charm, using the
\slj{3}{2}{0} model of $q\bar{q}$ production first applied above charm 
threshold by the Orsay group\cite{LeYaouanc:1977ux}.  They did not carry
out a coupled-channel analysis, and so did not determine the
composition of the physical states, but their estimates of open-charm 
decay rates can be read against ours as a rough assessment of model 
dependence.

The long-standing expectation that the 1\slj{3}{3}{2} and 1\slj{1}{3}{2} levels 
would be narrow followed from the presumption that these unnatural 
parity states should lie between the $D\bar{D}$ and $D\bar{D}^{*}$ 
thresholds, and could not decay into open charm. At $3872\mev$, both 
states can decay into $D^{0}\bar{D}^{*0}$, but the partial widths 
 are 
quite small. 
We show the variation of the 1\slj{3}{3}{2} partial width with mass in
the top right panel of Figure~\ref{figure:OCdecays}; over the region 
of interest, it does not threaten the Belle bound, 
$\Gamma(X(3872))<2.3\mev$. The range of values is quite similar to 
the range estimated for $\Gamma(1\slj{3}{3}{2} \to \pi\pi\jpsi)$, so 
we expect roughly comparable branching fractions for decays into 
$D^{0}\bar{D}^{*0}$ and $\pi^{+}\pi^{-}\jpsi$. If $X(3872)$ does turn 
out to be the 1\slj{3}{3}{2} level, we expect $M(1\slj{1}{3}{2}) = 
3880\mev$ and $\Gamma(1\slj{1}{3}{2} \to D^{0}\bar{D}^{*0}) \approx 
1.7\mev$.

The natural-parity 1\slj{3}{3}{3} state can decay into 
$D\bar{D}$, but its $f$-wave decay is suppressed by the centrifugal 
barrier factor, so the partial width is less than $1\mev$ at a mass 
of $3872\mev$. Although estimates of the hadronic cascade transitions 
are uncertain, the numbers in hand lead us to expect 
$\Gamma(1\slj{3}{3}{3} \to \pi^{+}\pi^{-}\jpsi) \ltap \cfrac{1}{4} 
\Gamma(1\slj{3}{3}{3} \to D\bar{D})$, whereas $\Gamma(1\slj{3}{3}{3} 
\to \gamma\chi_{c2}) \approx \cfrac{1}{3}\Gamma(1\slj{3}{3}{3} \to 
D\bar{D})$, if $X(3872)$ is identified as 1\slj{3}{3}{3}.
The variation of $\Gamma(1\slj{3}{3}{3} \to 
D\bar{D})$ with mass is shown in the 
middle left panel of Figure~\ref{figure:OCdecays}. Note that if 
1\slj{3}{3}{3} is not to be identified with $X(3872)$, \textit{it may still 
be discovered as a narrow $D\bar{D}$ resonance,} up to a mass of about 
$4000\mev$.

In their study of $B^{+} \to K^{+} \psi(3770)$ decays, the Belle 
Collaboration\cite{Abe:2003zv} has set 90\% CL upper limits on the 
transition $B^{+} \to K^{+} X(3872)$, followed by $X(3872) \to 
D\bar{D}$. Their limits imply that
\begin{eqnarray}
    \mathcal{B}(X(3872) \to D^{0}\bar{D}^{0}) & \ltap & 
    4\mathcal{B}(X \to \pi^{+}\pi^{-}\jpsi)\; , \nonumber \\[-6pt]
     & & 
    \label{eq:Xddbar}  \\[-6pt]
    \mathcal{B}(X(3872) \to D^{+}D^{-})  & \ltap & 
    3\mathcal{B}(X \to \pi^{+}\pi^{-}\jpsi)\; . \nonumber 
\end{eqnarray}
This constraint is already intriguingly close to the level at which we 
would expect to see $1\slj{3}{3}{3} \to D\bar{D}$.

The constraint on the total width of $X(3872)$ raises more of a 
challenge for the 2\slj{1}{2}{1} candidate, whose
$s$-wave decay to $D^{0}\bar{D}^{*0}$ rises dramatically from 
threshold, 
as shown in the middle right panel of Figure~\ref{figure:OCdecays}. 
Within the current uncertainty ($3871.7 \pm 0.6\mev$) in the mass of 
$X$, the issue cannot be settled, but the 2\slj{1}{2}{1} 
interpretation is viable only if $X$ lies below $D^{0}\bar{D}^{*0}$ 
threshold. If a light 2\slj{1}{2}{1} does turn out to be $X(3872)$, 
then its 2\slj{3}{2}{J} partners should lie nearby. In that case, 
they should be visible as relatively narrow charm-anticharm 
resonances. At $3872\mev$, we estimate  $\Gamma(2\slj{3}{2}{1}\to D\bar{D}^{*}) 
\approx 21\mev$ and $\Gamma(2\slj{3}{2}{2}\to D\bar{D}) \approx 
3\mev$. The bottom left panel in Figure~\ref{figure:OCdecays} shows 
that the 2\slj{3}{2}{2} level remains relatively narrow up to the 
opening of the $D^{*}\bar{D}^{*}$ threshold.

I point out one more candidate for a narrow resonance of charmed 
mesons: The 1\slj{3}{4}{4} level remains narrow 
($\Gamma(1\slj{3}{4}{4} \to \hbox{charm}) \ltap 
5\mev$) up to the 
$D^{*}\bar{D}^{*}$ threshold, as 
illustrated in the bottom right panel of Figure~\ref{figure:OCdecays}. Its allowed decays into $D\bar{D}$ and 
$D\bar{D}^{*}$ are inhibited by $\ell = 4$ barrier factors, whereas 
the $D^{*}\bar{D}^{*}$ channel is reached by $\ell = 2$.
%
\section{Following up the Discovery of $X(3872)$}
On the experimental front, the first order of business is to establish
the nature of $X(3872)$.  Determining the spin-parity of $X$ will
winnow the field of candidates.  The charmonium interpretation and its
prominent rivals require that $X(3872)$ be a neutral isoscalar.  Are
there charged partners?  A search for $X(3872) \to \pi^{0}\pi^{0}\jpsi$
will be highly informative.  As Barnes \& Godfrey\cite{Barnes:2003vb}
have remarked, observing a significant $\pi^{0}\pi^{0}\jpsi$ signal
establishes that $X$ is odd under charge conjugation.  Voloshin has
commented\cite{VoloshinPC} that the ratio $\mathcal{R}_{0} \equiv
\Gamma(X\to \pi^{0}\pi^{0}\jpsi)/\Gamma(X \to \pi^{+}\pi^{-}\jpsi)$
measures the dipion isospin.  Writing $\Gamma_{I}\equiv \Gamma(X \to
(\pi^{+}\pi^{-})_{I}\jpsi)$, we see that $\mathcal{R}_{0} =
\cfrac{1}{2}/(1 + \Gamma_{1}/\Gamma_{0})$, up to kinematic corrections.
Deviations from $\mathcal{R}_{0}=\cfrac{1}{2}$ signal the
isospin-violating decay of an isoscalar, or the isospin-conserving
decay of an isovector.  Radiative decay rates and the prompt (as
opposed to $B$-decay) production fraction will provide important
guidance.  Other diagnostics of a general nature have been discussed
in Refs.\cite{Close:2003mb,Barnes:2003vb,Pakvasa:2003ea,Close:2003sg}.

Within the charmonium framework, $X(3872)$ is most naturally 
interpreted as the 1\slj{3}{3}{2} or 1\slj{3}{3}{3} level, both of 
which have allowed decays into $\pi\pi\jpsi$. The 
$2^{--}$ 1\slj{3}{3}{2} state is forbidden by parity conservation to 
decay into $D\bar{D}$ but has a modest $D^{0}\bar{D}^{*0}$ partial width 
for masses near $3872\mev$. Although the uncertain $\pi\pi\jpsi$ 
partial width makes it difficult to estimate relative branching 
ratios, the decay $X(3872) \to \chi_{c1}\,\gamma(344)$ should show itself 
if $X$ is indeed 1\slj{3}{3}{2}. The $\chi_{c2}\,\gamma(303)$ line 
should be seen with about $\cfrac{1}{4}$ the strength of 
$\chi_{c1}\,\gamma(344)$.  In our coupled-channel calculation, the 
1\slj{3}{3}{2} mass is about $41\mev$ lower than the observed 
$3872\mev$. In contrast, the computed 1\slj{3}{3}{3} mass is quite 
close to $3872\mev$, and 1\slj{3}{3}{3} does not have an E1 
transition to $\chi_{c1}\,\gamma(344)$. The dominant decay of the 
$3^{--}$ 1\slj{3}{3}{3} state should be into $D\bar{D}$; a small 
branching fraction for the $\pi\pi\jpsi$ discovery 
mode would imply a large production rate. One radiative 
transition should be observable, with $\Gamma(X(3872) \to 
\chi_{c2}\,\gamma(303)) \gtap \Gamma(X(3872) \to \pi^{+}\pi^{-}\jpsi)$. 
I underscore the importance of searching for the $\chi_{c1}\,\gamma(344)$ 
and $\chi_{c2}\,\gamma(303)$ lines.
    
Beyond pinning down the character of $X(3872)$, experiments can search 
for additional narrow charmonium states in radiative and hadronic 
transitions to lower-lying $c\bar{c}$ levels, as we emphasized in 
Ref.\cite{Eichten:2002qv}, and in neutral combinations of charmed 
mesons and anticharmed mesons. The coupled-channel analysis presented 
in our most recent paper\cite{Eichten:2004uh} sets up specific targets.

More broadly, it is worth reminding ourselves that a search for structure in channels including $\jpsi$ and any readily detectable hadron, including $\jpsi+ \pi^\pm, \eta, K^\pm, K_s, p, \Lambda, \ldots$ may be rewarding. Some of these combinations are predicted to be there, and others just might exist. We would feel foolish if we failed to look.

On the theoretical front, we need a more complete understanding of the
production of the charmonium states in $B$ decays and by direct
hadronic production, including the influence of open-charm channels.
Understanding of the production mechanisms for molecular charm or
$c\bar{c}g$ hybrid states is much more primitive.  The hybrid-meson hypothesis in particular needs some specific predictions and a decision tree to test the interpretation. We need to improve
the theoretical understanding of hadronic cascades among charmonium
states, including the influence of open-charm channels.  The comparison
of charmonium transitions with their upsilon counterparts should be
informative.  The analysis we have carried out can be extended to the
$b\bar{b}$ system, where it may be possible to see discrete
threshold-region states in direct hadronic production.  Because the
Cornell coupled-channel model is only an approximation to QCD, it would
be highly desirable to compare its predictions with those of a
coupled-channel analysis of the \slj{3}{2}{0} model of quark pair
production. Ultimately, extending lattice QCD calculations into the 
flavor-threshold region should give a firmer basis for predictions.

In addition to the $1\slj{1}{2}{1}\;h_{c}$, the now-established 
$2\slj{1}{1}{0}\;\eta_{c}^{\prime}$, and the long-sought 
$1\slj{1}{3}{2}\;\eta_{c2}$ and $1\slj{3}{3}{2}\;\psi_{2}$ states, 
discrete charmonium levels are to be found as narrow charm-anticharm 
structures in the flavor-threshold region. The most likely candidates 
correspond to the $1\slj{3}{3}{3}$, $2\slj{3}{2}{2}$, and 
$1\slj{3}{4}{4}$ levels. If $X(3872)$ is indeed a charmonium 
state---the \slj{3}{3}{2} and \slj{3}{3}{3} assignments seem most 
promising---then identifying that state anchors the mass scale. If 
$X(3872)$ is not charmonium, then all the charmonium levels remain to 
be discovered.
Finding these states---and establishing their 
masses, widths, and production rates---will lead us into new terrain.

\section{The Next Wave: Mesons with Beauty and Charm}
\begin{figure}[tb]
   \centerline{ \BoxedEPSF{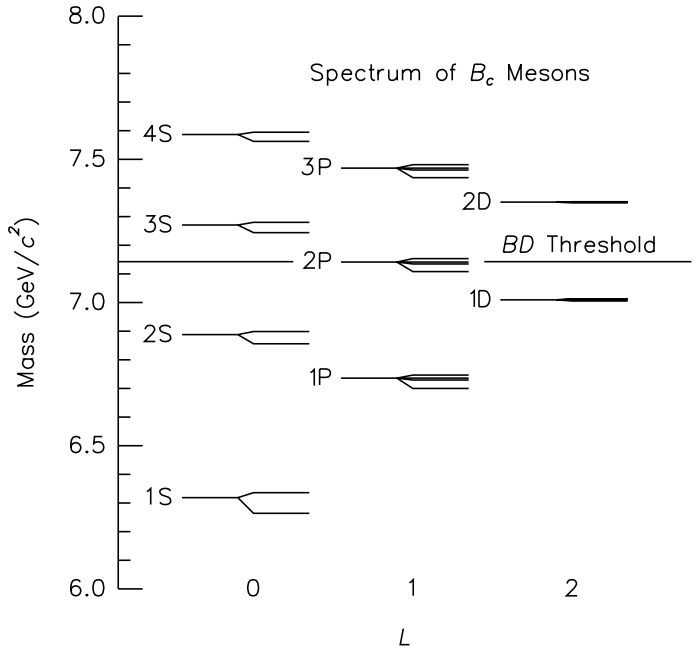 scaled 600}}%
\caption{The spectrum of $b\bar{c}$ mesons, from Ref.\cite{Eichten:1994gt}.}
\label{fig:bcbar}
\end{figure}

Before closing, I want to make a few remarks about the $b\bar{c}$ system, which I hope will be studied in some detail in Run II of the Tevatron Collider, and later in ATLAS and CMS at the Large Hadron Collider. Knowing the interquark potential and the masses of the $b$- and $c$-quarks, we can readily compute the spectrum of $b\bar{c}$ bound states. This has now been done by many people, who find results similar to those displayed in Figure~\ref{fig:bcbar}, taken from my work with Eichten\cite{Eichten:1994gt}. 
Below the $B\bar{D}$ threshold, we expect the 1S and 2S doublets as well as the 1P and 1D quartets, and perhaps some of the 2P states. Because the $b\bar{c}$ mesons have both beauty and charm, {all the excited states make radiative or hadronic cascades to the ground state.} There are no annihilations into gluons such as we encounter in the charmonium or upsilon families. Among the easily identified decays of $B_c^+$ should be $\jpsi\,\pi^+$, $\jpsi\,a_1^+$, and $\jpsi\,\ell^+\nu$. By considering a universe of reasonable quarkonium potentials, we estimated that the ground-state mass would lie within $20\mev$ of $6258\mev$.

We have a number of reasons to want to explore this third system and map out its spectrum. First, it is a wonderful experimental challenge: it will be a remarkable experimental tour-de-force to establish the $B_c$ ground state and some of the transitions that lead to it. Second, the $b\bar{c}$ family is intermediate between a heavy-heavy system and a heavy-light system; it should display attributes of both, and theoretical techniques developed for the limiting cases may find interesting challenges here. Third, $B_c$ and its excited states should be sensitive to relativistic effects and configuration mixing. It is worth noting that the $c$-quark in the $B_c$ has a much higher velocity than its counterpart in charmonium. Finally, the rich pattern of weak decays: $b$-decay, $c$-decay, and $b\bar{c}$ annihilation, should have much to teach us about the interplay of weak and strong interactions.

In 1998, the CDF Collaboration observed\cite{Abe:1998wi} the semileptonic decay $B_c \to \jpsi\,\ell\nu$, and inferred a ground-state mass $M_{B_c} =  6.40 \pm 0.39\hbox{ (stat.)} \pm  0.13\hbox{ (sys.)}\gev$. No nonleptonic decay has yet been established. Using the semileptonic sample, CDF measured a lifetime $\tau(B_c) = 0.46^{+0.18}_{-0.16}\hbox{ (stat.)} \pm 0.03\hbox{ (syst.)}\hbox{ ps}$, consistent with theoretical expectations\cite{Beneke:1996xe}.

\begin{figure}[tb]
\centerline{\BoxedEPSF{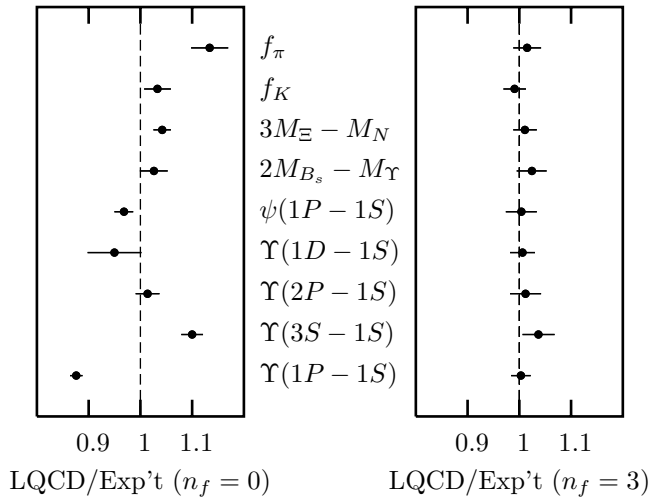 scaled 1000}}
\caption{Lattice QCD results divided by experimental results for nine different
quantities, without (left panel) and with (right panel) quark vacuum polarization, from Ref.\cite{Davies:2003ik}. The top three results are from simulations with lattice spacings of $\cfrac{1}{11}$ and $\cfrac{1}{8}\fm$; all others are from $\cfrac{1}{8}$-fm simulations.}
\label{fig:HPQCD}
\end{figure}

The High-Precision QCD collaboration\cite{Davies:2003ik} has recently reported important progress in the inclusion of dynamical fermions in lattice calculations of quarkonium observables. Their test of lattice predictions consists in tuning the bare $u$- and $d$-quark masses (set equal), the bare~$s$-, $c$- and
$b$-masses, and a proxy for the bare QCD coupling, to reproduce $M_\pi^2$, $2M_K^2-M_\pi^2$, $M_{D_s}$, $M_\Upsilon$, and $M_{\Upsilon^\prime} - M_\Upsilon$.
Having tuned all free parameters, they computed nine other observables. The outcome of their calculations is summarized in Figure~\ref{fig:HPQCD}, which shows the ratio
of calculation to measurement for the pion and kaon decay
constants, a baryon mass splitting, the $B_s$--$\Upsilon$ mass difference, and mass
differences between various $c\bar{c}$ and $b\bar{b}$ states. The left panel shows
ratios from quenched QCD simulations without quark vacuum polarization.  These
results deviate from experiment by as much as~10--15\%. The right panel shows results from unquenched QCD simulations that include realistic vacuum polarization. With no free parameters, the unquenched calculations reproduce experiment to within systematic and statistical errors of~3\% or less.

At the recent Aspen Winter Physics Conference, Andreas Kronfeld reported new predictions\cite{ASKAspen} for the mass of the $b\bar{c}$ ground state on behalf of a Glasgow--Fermilab subset of the HPQCD Collaboration. Using quarkonium as a baseline, they quote a preliminary value, $M_{B_c} = 6307 \pm 2 ^{+0}_{-10}\mev$; using a heavy-light baseline, they find
$M_{B_c} = 6253 \pm 17 ^{+30\mathrm{-}50}_{-0}\mev$. They are making final studies of the sensitivity to lattice spacing and the sea-quark mass, and expect to present a final result soon.

The new lattice result joins older potential-model estimates as an attractive target for experiment, with the significant added benefit that it includes the full richness of strong-interaction dynamics. We look forward to an early experimental verdict!

\section{Outlook}
The discovery of the narrow state $X(3872) \to \pi^+\pi^-\jpsi$ gives quarkonium physics a rich and lively puzzle. We do not yet know what this state is. If the most conventional interpretation as a charmonium state---most plausibly, the 1\slj{3}{3}{2} or 1\slj{3}{3}{3} level---is confirmed, we will learn important lessons about the influence of open-charm states on $c\bar{c}$ levels. Should the charmonium interpretation not prevail, perhaps $X(3872)$ will herald an entirely new spectroscopy. In either event, several new charmonium states remain to be discovered through their radiative decays or hadronic transitions to lower $c\bar{c}$ levels. Another set of $c\bar{c}$ states promise to be observable as narrow structures that decay into pairs of charmed mesons. In time, comparing what we learn from this new exploration of the charmonium spectrum with analogous states in the $b\bar{b}$ family will be rewarding. We have seen the first experimental evidence for a third, and rather exotic, quarkonium family, the mesons with beauty and charm of the $b\bar{c}$ series. A precise determination of the ground-state $B_c$ mass will test the state of the lattice QCD art, and mapping the spectrum will enhance our understanding of quarkonium systems. The weak decays of $B_c$ should be highly instructive. For all three quarkonium families, we need to improve our understanding of hadronic cascades. Beyond spectroscopy, we look forward to new insights about the production of quarkonium states in $B$ decays and hard scattering. I expect continued good fun from the interchange between theory and experiment!
\section{Acknowledgements}
It is a pleasure to thank the organizers of the XVIII Rencontres de Physique de la Vall\'{e}e d'Aoste for the excellent scientific and human atmosphere in La Thuile.
Fermilab is operated by
Universities Research Association Inc.\ under Contract No.\
DE-AC02-76CH03000 with the U.S.\ Department of Energy.
I thank Estia Eichten and Ken Lane for a stimulating collaboration on the matters reported here.
\bibliography{Quigg04}

\begin{thebibliography}{49}
\expandafter\ifx\csname natexlab\endcsname\relax\def\natexlab#1{#1}\fi
\expandafter\ifx\csname bibnamefont\endcsname\relax
  \def\bibnamefont#1{#1}\fi
\expandafter\ifx\csname bibfnamefont\endcsname\relax
  \def\bibfnamefont#1{#1}\fi
\expandafter\ifx\csname citenamefont\endcsname\relax
  \def\citenamefont#1{#1}\fi
\expandafter\ifx\csname url\endcsname\relax
  \def\url#1{\texttt{#1}}\fi
\expandafter\ifx\csname urlprefix\endcsname\relax\def\urlprefix{URL }\fi
\providecommand{\bibinfo}[2]{#2}
\providecommand{\eprint}[2][]{\url{#2}}

\bibitem{Choi:2002na}
\bibinfo{author}{\bibfnamefont{S.~K.} \bibnamefont{Choi}} \bibnamefont{et~al.}
  (\bibinfo{collaboration}{Belle}), \bibinfo{journal}{Phys. Rev. Lett.}
  \textbf{\bibinfo{volume}{89}}, \bibinfo{pages}{102001}
  (\bibinfo{year}{2002}), \bibinfo{note}{erratum: ibid.\ \textbf{89,} 129901
  (2002)}, \eprint{hep-ex/0206002}.

\bibitem{Asner:2003wv}
\bibinfo{author}{\bibfnamefont{D.~M.} \bibnamefont{Asner}}
  (\bibinfo{collaboration}{CLEO}) (\bibinfo{year}{2003}),
  \eprint{hep-ex/0312058}.

\bibitem{Wagner:2003qb}
\bibinfo{author}{\bibfnamefont{G.}~\bibnamefont{Wagner}}
  (\bibinfo{collaboration}{BABAR}) (\bibinfo{year}{2003}),
  \eprint{hep-ex/0305083}.

\bibitem{Abe:2003ja}
\bibinfo{author}{\bibfnamefont{K.}~\bibnamefont{Abe}} \bibnamefont{et~al.}
  (\bibinfo{collaboration}{Belle}) (\bibinfo{year}{2003}),
  \eprint{hep-ex/0306015}.

\bibitem{Skwarnicki:2003wn}
\bibinfo{author}{\bibfnamefont{T.}~\bibnamefont{Skwarnicki}}
  (\bibinfo{year}{2003}), \eprint{hep-ph/0311243}.

\bibitem{Aubert:2003fg}
\bibinfo{author}{\bibfnamefont{B.}~\bibnamefont{Aubert}} \bibnamefont{et~al.}
  (\bibinfo{collaboration}{BABAR}), \bibinfo{journal}{Phys. Rev. Lett.}
  \textbf{\bibinfo{volume}{90}}, \bibinfo{pages}{242001}
  (\bibinfo{year}{2003}), \eprint{hep-ex/0304021}.

\bibitem{Besson:2003cp}
\bibinfo{author}{\bibfnamefont{D.}~\bibnamefont{Besson}} \bibnamefont{et~al.}
  (\bibinfo{collaboration}{CLEO}), \bibinfo{journal}{Phys. Rev.}
  \textbf{\bibinfo{volume}{D68}}, \bibinfo{pages}{032002}
  (\bibinfo{year}{2003}), \eprint{hep-ex/0305100}.

\bibitem{Abe:2003jk}
\bibinfo{author}{\bibfnamefont{K.}~\bibnamefont{Abe}} \bibnamefont{et~al.},
  \bibinfo{journal}{Phys. Rev. Lett.} \textbf{\bibinfo{volume}{92}},
  \bibinfo{pages}{012002} (\bibinfo{year}{2004}), \eprint{hep-ex/0307052}.

\bibitem{Choi:2003ue}
\bibinfo{author}{\bibfnamefont{S.~K.} \bibnamefont{Choi}} \bibnamefont{et~al.}
  (\bibinfo{collaboration}{Belle}), \bibinfo{journal}{Phys. Rev. Lett.}
  \textbf{\bibinfo{volume}{91}}, \bibinfo{pages}{262001}
  (\bibinfo{year}{2003}), \eprint{hep-ex/0309032}.

\bibitem{Acosta:2003zx}
\bibinfo{author}{\bibfnamefont{D.}~\bibnamefont{Acosta}} \bibnamefont{et~al.}
  (\bibinfo{collaboration}{CDF II Collaboration}) (\bibinfo{year}{2003}),
  \eprint{hep-ex/0312021}.

\bibitem{Dzerotemp}
\bibinfo{author}{\bibnamefont{{{D\O} Collaboration}}} (\bibinfo{year}{2004}),
  \bibinfo{note}{{{D\O} Note 4334}}, \urlprefix\url{http://www-d0.fnal.gov
  /Run2Physics/ckm/Moriond\_2003/X\_conf{\_}note\_v9.ps}.

\bibitem{JesikLT}
\bibinfo{author}{\bibfnamefont{R.}~\bibnamefont{Jesik}} (\bibinfo{year}{2004}),
  \bibinfo{note}{\textit{This Volume}},
  \urlprefix\url{http://www.pi.infn.it/lathuile/
  2004/talks/contributi/jesik.pdf}.

\bibitem{Bardeen:2003kt}
\bibinfo{author}{\bibfnamefont{W.~A.} \bibnamefont{Bardeen}},
  \bibinfo{author}{\bibfnamefont{E.~J.} \bibnamefont{Eichten}},
  \bibnamefont{and} \bibinfo{author}{\bibfnamefont{C.~T.} \bibnamefont{Hill}},
  \bibinfo{journal}{Phys. Rev.} \textbf{\bibinfo{volume}{D68}},
  \bibinfo{pages}{054024} (\bibinfo{year}{2003}), \eprint{hep-ph/0305049}.

\bibitem{VaiaLT}
\bibinfo{author}{\bibfnamefont{V.}~\bibnamefont{Papadimitriou}}
  (\bibinfo{year}{2004}), \bibinfo{note}{\textit{This Volume}},
  \urlprefix\url{http://www.pi.infn.it/
  lathuile/2004/talks/contributi/papadimitriou.pdf}.

\bibitem{Eichten:2002qv}
\bibinfo{author}{\bibfnamefont{E.~J.} \bibnamefont{Eichten}},
  \bibinfo{author}{\bibfnamefont{K.}~\bibnamefont{Lane}}, \bibnamefont{and}
  \bibinfo{author}{\bibfnamefont{C.}~\bibnamefont{Quigg}},
  \bibinfo{journal}{Phys. Rev. Lett.} \textbf{\bibinfo{volume}{89}},
  \bibinfo{pages}{162002} (\bibinfo{year}{2002}), \eprint{hep-ph/0206018}.

\bibitem{Ko:1997rn}
\bibinfo{author}{\bibfnamefont{P.}~\bibnamefont{Ko}},
  \bibinfo{author}{\bibfnamefont{J.}~\bibnamefont{Lee}}, \bibnamefont{and}
  \bibinfo{author}{\bibfnamefont{H.~S.} \bibnamefont{Song}},
  \bibinfo{journal}{Phys. Lett.} \textbf{\bibinfo{volume}{B395}},
  \bibinfo{pages}{107} (\bibinfo{year}{1997}),
  \eprint[http://arXiv.org/abs]{hep-ph/9701235}.

\bibitem{Suzuki:2002sq}
\bibinfo{author}{\bibfnamefont{M.}~\bibnamefont{Suzuki}}
  (\bibinfo{year}{2002}), \eprint[http://arXiv.org/abs]{hep-ph/0204043}.

\bibitem{Bodwin:1992qr}
\bibinfo{author}{\bibfnamefont{G.~T.} \bibnamefont{Bodwin}},
  \bibinfo{author}{\bibfnamefont{E.}~\bibnamefont{Braaten}},
  \bibinfo{author}{\bibfnamefont{T.~C.} \bibnamefont{Yuan}}, \bibnamefont{and}
  \bibinfo{author}{\bibfnamefont{G.~P.} \bibnamefont{Lepage}},
  \bibinfo{journal}{Phys. Rev.} \textbf{\bibinfo{volume}{D46}},
  \bibinfo{pages}{3703} (\bibinfo{year}{1992}),
  \eprint[http://arXiv.org/abs]{hep-ph/9208254}.

\bibitem{Ko:1996iv}
\bibinfo{author}{\bibfnamefont{P.}~\bibnamefont{Ko}},
  \bibinfo{author}{\bibfnamefont{J.}~\bibnamefont{Lee}}, \bibnamefont{and}
  \bibinfo{author}{\bibfnamefont{H.~S.} \bibnamefont{Song}},
  \bibinfo{journal}{Phys. Rev.} \textbf{\bibinfo{volume}{D53}},
  \bibinfo{pages}{1409} (\bibinfo{year}{1996}),
  \eprint[http://arXiv.org/abs]{hep-ph/9510202}.

\bibitem{Yuan:1997we}
\bibinfo{author}{\bibfnamefont{F.}~\bibnamefont{Yuan}},
  \bibinfo{author}{\bibfnamefont{C.-F.} \bibnamefont{Qiao}}, \bibnamefont{and}
  \bibinfo{author}{\bibfnamefont{K.-T.} \bibnamefont{Chao}},
  \bibinfo{journal}{Phys. Rev.} \textbf{\bibinfo{volume}{D56}},
  \bibinfo{pages}{329} (\bibinfo{year}{1997}),
  \eprint[http://arXiv.org/abs]{hep-ph/9701250}.

\bibitem{ZhuLT}
\bibinfo{author}{\bibfnamefont{Y.-S.} \bibnamefont{Zhu}}
  (\bibinfo{year}{2004}), \bibinfo{note}{\textit{This Volume}},
  \urlprefix\url{http://www.pi.infn.it/
  lathuile/2004/talks/contributi/zhu.pdf}.

\bibitem{Bai:2003hv}
\bibinfo{author}{\bibfnamefont{J.~Z.} \bibnamefont{Bai}} \bibnamefont{et~al.}
  (\bibinfo{collaboration}{BES}) (\bibinfo{year}{2003}),
  \eprint{hep-ex/0307028}.

\bibitem{ChoiLL}
\bibinfo{author}{\bibfnamefont{S.-K.} \bibnamefont{Choi}}
  (\bibinfo{year}{2004}), \bibinfo{note}{\textrm{Lake Louise Winter
  Institute}}, \urlprefix\url{http://
  www.phys.hawaii.edu/\~{}solsen/x3872/lake\_louise\_2004.pdf}.

\bibitem{JDJhouches}
\bibinfo{author}{\bibfnamefont{J.~D.} \bibnamefont{Jackson}}, in
  \emph{\bibinfo{booktitle}{High Energy Physics: Les Houches 1965}}, edited by
  \bibinfo{editor}{\bibfnamefont{C.}~\bibnamefont{{De Witt}}} \bibnamefont{and}
  \bibinfo{editor}{\bibfnamefont{M.}~\bibnamefont{{Jacob}}}
  (\bibinfo{publisher}{Gordon \& Breach}, \bibinfo{year}{1965}), pp.
  \bibinfo{pages}{325--365}, \bibinfo{note}{and private communication, 2003}.

\bibitem{Pakvasa:2003ea}
\bibinfo{author}{\bibfnamefont{S.}~\bibnamefont{Pakvasa}} \bibnamefont{and}
  \bibinfo{author}{\bibfnamefont{M.}~\bibnamefont{Suzuki}},
  \bibinfo{journal}{Phys. Lett.} \textbf{\bibinfo{volume}{B579}},
  \bibinfo{pages}{67} (\bibinfo{year}{2004}), \eprint{hep-ph/0309294}.

\bibitem{Yuan:2003yz}
\bibinfo{author}{\bibfnamefont{C.~Z.} \bibnamefont{Yuan}},
  \bibinfo{author}{\bibfnamefont{X.~H.} \bibnamefont{Mo}}, \bibnamefont{and}
  \bibinfo{author}{\bibfnamefont{P.}~\bibnamefont{Wang}},
  \bibinfo{journal}{Phys. Lett.} \textbf{\bibinfo{volume}{B579}},
  \bibinfo{pages}{74} (\bibinfo{year}{2004}), \eprint{hep-ph/0310261}.

\bibitem{Tornqvist:2004qy}
\bibinfo{author}{\bibfnamefont{N.~A.} \bibnamefont{T\"{o}rnqvist}}
  (\bibinfo{year}{2004}), \bibinfo{note}{(supersedes \texttt{hep-ph/0308277})},
  \eprint{hep-ph/0402237}.

\bibitem{Voloshin:2003nt}
\bibinfo{author}{\bibfnamefont{M.~B.} \bibnamefont{Voloshin}},
  \bibinfo{journal}{Phys. Lett.} \textbf{\bibinfo{volume}{B579}},
  \bibinfo{pages}{316} (\bibinfo{year}{2004}), \eprint{hep-ph/0309307}.

\bibitem{Wong:2003xk}
\bibinfo{author}{\bibfnamefont{C.-Y.} \bibnamefont{Wong}}
  (\bibinfo{year}{2003}), \eprint{hep-ph/0311088}.

\bibitem{Swanson:2003tb}
\bibinfo{author}{\bibfnamefont{E.~S.} \bibnamefont{Swanson}}
  (\bibinfo{year}{2003}), \eprint{hep-ph/0311229}.

\bibitem{Abe:2003zv}
\bibinfo{author}{\bibfnamefont{R.}~\bibnamefont{Chistov}} \bibnamefont{et~al.}
  (\bibinfo{collaboration}{Belle}) (\bibinfo{year}{2003}),
  \eprint{hep-ex/0307061}.

\bibitem{Braaten:2003he}
\bibinfo{author}{\bibfnamefont{E.}~\bibnamefont{Braaten}} \bibnamefont{and}
  \bibinfo{author}{\bibfnamefont{M.}~\bibnamefont{Kusunoki}}
  (\bibinfo{year}{2003}), \eprint{hep-ph/0311147}.

\bibitem{Braaten:2004rn}
\bibinfo{author}{\bibfnamefont{E.}~\bibnamefont{Braaten}} \bibnamefont{and}
  \bibinfo{author}{\bibfnamefont{M.}~\bibnamefont{Kusunoki}}
  (\bibinfo{year}{2004}), \eprint{hep-ph/0402177}.

\bibitem{Close:2003mb}
\bibinfo{author}{\bibfnamefont{F.~E.} \bibnamefont{Close}} \bibnamefont{and}
  \bibinfo{author}{\bibfnamefont{S.}~\bibnamefont{Godfrey}},
  \bibinfo{journal}{Phys. Lett.} \textbf{\bibinfo{volume}{B574}},
  \bibinfo{pages}{210} (\bibinfo{year}{2003}), \eprint{hep-ph/0305285}.

\bibitem{Aubert:2004fc}
\bibinfo{author}{\bibfnamefont{B.}~\bibnamefont{Aubert}} \bibnamefont{et~al.}
  (\bibinfo{collaboration}{BABAR}) (\bibinfo{year}{2004}),
  \eprint{hep-ex/0402025}.

\bibitem{Eichten:1978tg}
\bibinfo{author}{\bibfnamefont{E.}~\bibnamefont{Eichten}},
  \bibinfo{author}{\bibfnamefont{K.}~\bibnamefont{Gottfried}},
  \bibinfo{author}{\bibfnamefont{T.}~\bibnamefont{Kinoshita}},
  \bibinfo{author}{\bibfnamefont{K.~D.} \bibnamefont{Lane}}, \bibnamefont{and}
  \bibinfo{author}{\bibfnamefont{T.-M.} \bibnamefont{Yan}},
  \bibinfo{journal}{Phys. Rev.} \textbf{\bibinfo{volume}{D17}},
  \bibinfo{pages}{3090} (\bibinfo{year}{1978}), \bibinfo{note}{[Erratum-ibid.\
  D {\bf 21}, 313 (1980)].}

\bibitem{Eichten:1980ms}
\bibinfo{author}{\bibfnamefont{E.}~\bibnamefont{Eichten}},
  \bibinfo{author}{\bibfnamefont{K.}~\bibnamefont{Gottfried}},
  \bibinfo{author}{\bibfnamefont{T.}~\bibnamefont{Kinoshita}},
  \bibinfo{author}{\bibfnamefont{K.~D.} \bibnamefont{Lane}}, \bibnamefont{and}
  \bibinfo{author}{\bibfnamefont{T.-M.} \bibnamefont{Yan}},
  \bibinfo{journal}{Phys. Rev.} \textbf{\bibinfo{volume}{D21}},
  \bibinfo{pages}{203} (\bibinfo{year}{1980}).

\bibitem{Eichten:2004uh}
\bibinfo{author}{\bibfnamefont{E.~J.} \bibnamefont{Eichten}},
  \bibinfo{author}{\bibfnamefont{K.}~\bibnamefont{Lane}}, \bibnamefont{and}
  \bibinfo{author}{\bibfnamefont{C.}~\bibnamefont{Quigg}}
  (\bibinfo{year}{2004}), \eprint{hep-ph/0401210}.

\bibitem{Recksiegel:2003fm}
\bibinfo{author}{\bibfnamefont{S.}~\bibnamefont{Recksiegel}} \bibnamefont{and}
  \bibinfo{author}{\bibfnamefont{Y.}~\bibnamefont{Sumino}},
  \bibinfo{journal}{Phys. Lett.} \textbf{\bibinfo{volume}{B578}},
  \bibinfo{pages}{369} (\bibinfo{year}{2004}), \eprint{hep-ph/0305178}.

\bibitem{PDBook}
\bibinfo{author}{\bibfnamefont{K.}~\bibnamefont{{Hagiwara}}}
  \bibnamefont{et~al.} (\bibinfo{collaboration}{Particle Data Group}),
  \bibinfo{journal}{{Phys. Rev. D}} \textbf{\bibinfo{volume}{66}},
  \bibinfo{pages}{010001} (\bibinfo{year}{2002}), \bibinfo{note}{and 2003
  off-year partial update}, \urlprefix\url{http://pdg.lbl.gov}.

\bibitem{Barnes:2003vb}
\bibinfo{author}{\bibfnamefont{T.}~\bibnamefont{Barnes}} \bibnamefont{and}
  \bibinfo{author}{\bibfnamefont{S.}~\bibnamefont{Godfrey}}
  (\bibinfo{year}{2003}), \eprint{hep-ph/0311162}.

\bibitem{LeYaouanc:1977ux}
\bibinfo{author}{\bibfnamefont{A.}~\bibnamefont{Le~Yaouanc}},
  \bibinfo{author}{\bibfnamefont{L.}~\bibnamefont{Oliver}},
  \bibinfo{author}{\bibfnamefont{O.}~\bibnamefont{P\`{e}ne}}, \bibnamefont{and}
  \bibinfo{author}{\bibfnamefont{J.~C.} \bibnamefont{Raynal}},
  \bibinfo{journal}{Phys. Lett.} \textbf{\bibinfo{volume}{B71}},
  \bibinfo{pages}{397} (\bibinfo{year}{1977}).

\bibitem{VoloshinPC}
\bibinfo{author}{\bibfnamefont{M.}~\bibnamefont{Voloshin}}
  (\bibinfo{year}{2003}), \bibinfo{note}{remarks at the International Workshop
  on Heavy Quarkonium, Fermilab, 20--23 September}.

\bibitem{Close:2003sg}
\bibinfo{author}{\bibfnamefont{F.~E.} \bibnamefont{Close}} \bibnamefont{and}
  \bibinfo{author}{\bibfnamefont{P.~R.} \bibnamefont{Page}},
  \bibinfo{journal}{Phys. Lett.} \textbf{\bibinfo{volume}{B578}},
  \bibinfo{pages}{119} (\bibinfo{year}{2004}), \eprint{hep-ph/0309253}.

\bibitem{Eichten:1994gt}
\bibinfo{author}{\bibfnamefont{E.~J.} \bibnamefont{Eichten}} \bibnamefont{and}
  \bibinfo{author}{\bibfnamefont{C.}~\bibnamefont{Quigg}},
  \bibinfo{journal}{Phys. Rev.} \textbf{\bibinfo{volume}{D49}},
  \bibinfo{pages}{5845} (\bibinfo{year}{1994}).

\bibitem{Abe:1998wi}
\bibinfo{author}{\bibfnamefont{F.}~\bibnamefont{Abe}} \bibnamefont{et~al.}
  (\bibinfo{collaboration}{CDF}), \bibinfo{journal}{Phys. Rev. Lett.}
  \textbf{\bibinfo{volume}{81}}, \bibinfo{pages}{2432} (\bibinfo{year}{1998}),
  \eprint{hep-ex/9805034}.

\bibitem{Beneke:1996xe}
\bibinfo{author}{\bibfnamefont{M.}~\bibnamefont{Beneke}} \bibnamefont{and}
  \bibinfo{author}{\bibfnamefont{G.}~\bibnamefont{Buchalla}},
  \bibinfo{journal}{Phys. Rev.} \textbf{\bibinfo{volume}{D53}},
  \bibinfo{pages}{4991} (\bibinfo{year}{1996}), \eprint{hep-ph/9601249}.

\bibitem{Davies:2003ik}
\bibinfo{author}{\bibfnamefont{C.~T.~H.} \bibnamefont{Davies}}
  \bibnamefont{et~al.} (\bibinfo{collaboration}{HPQCD}),
  \bibinfo{journal}{Phys. Rev. Lett.} \textbf{\bibinfo{volume}{92}},
  \bibinfo{pages}{022001} (\bibinfo{year}{2004}), \eprint{hep-lat/0304004}.

\bibitem{ASKAspen}
\bibinfo{author}{\bibfnamefont{A.}~\bibnamefont{Kronfeld}}
  (\bibinfo{year}{2004}), \bibinfo{note}{\textrm{Aspen Winter Particle Physics
  Conference}},
  \urlprefix\url{http://gate.hep.anl.gov/berger/Aspen04/Prog04/Kronfeld.pdf}.

\end{thebibliography}
\end{document}